\documentclass[aps,physrev,twocolumn,groupedaddress]{revtex4-2}
\usepackage{amsmath} 
\usepackage{graphicx}
\usepackage{hyperref}
\usepackage{circledsteps}

\usepackage[table]{xcolor}
\usepackage{cleveref}
\crefname{equation}{Eq.}{Eqs.}
\crefrangelabelformat{equation}{(#3#1#4--#5#2#6)}
\usepackage{bm}

\begin{document}

%\preprint{}

\title{Optimizing the interaction geometry of inverse Compton scattering x-ray sources}

\author{C.W. Sweers}
\email{c.w.sweers@tue.nl}
\affiliation{Department of Applied Physics, Eindhoven University of Technology, P.O. Box 513, 5600 MB Eindhoven, The Netherlands}
\author{O.J. Luiten}
\affiliation{Department of Applied Physics, Eindhoven University of Technology, P.O. Box 513, 5600 MB Eindhoven, The Netherlands}

\date{\today}

\begin{abstract}
Inverse Compton scattering (ICS) is a promising method for generating coherent and tunable x-rays in a compact setup. In this paper, we present a theoretical framework describing the output of an ICS x-ray source for arbitrary interaction angles between pulsed electron and laser beams, in the Thomson regime. This allows for analytic optimization of the x-ray beam properties by varying the parameters defining the geometry. In general, different x-ray applications require optimization of different x-ray beam properties, such as energy spread for x-ray spectroscopy and angular spread for x-ray scattering measurements. In this paper, we restrict ourselves to optimization of the x-ray brilliance, which is a comprehensive figure of merit for x-ray beam quality. The framework can be used, however, to optimize other x-ray properties. We investigate two specific ICS interaction geometries in particular: head-on scattering of a laser beam off an electron beam; and scattering of a laser beam off an electron beam in a co-propagating geometry, interacting under a grazing angle. For head-on scattering we show that a tightly focused, cylindrically symmetric laser pulse, which balances laser intensity and interaction time, optimizes the x-ray brilliance. For a co-propagating, grazing angle geometry, an elliptical focus of the laser pulse is required to mitigate the geometric reduction of the interaction time. We find that the latter geometry is especially useful for soft x-ray generation.
\end{abstract}

%\keywords{}

\maketitle
%%%%%%%%%%%%%%%%%%%%%%%%%%%%%%%%%%%%%%%%%%%%%%%%
%%%%%%%%%%%%%%%%%%%%%%%%%%%%%%%%%%%%%%%%%%%%%%%%
\section{Introduction}

\begin{figure*}
    \centering
     \includegraphics[width=\linewidth]{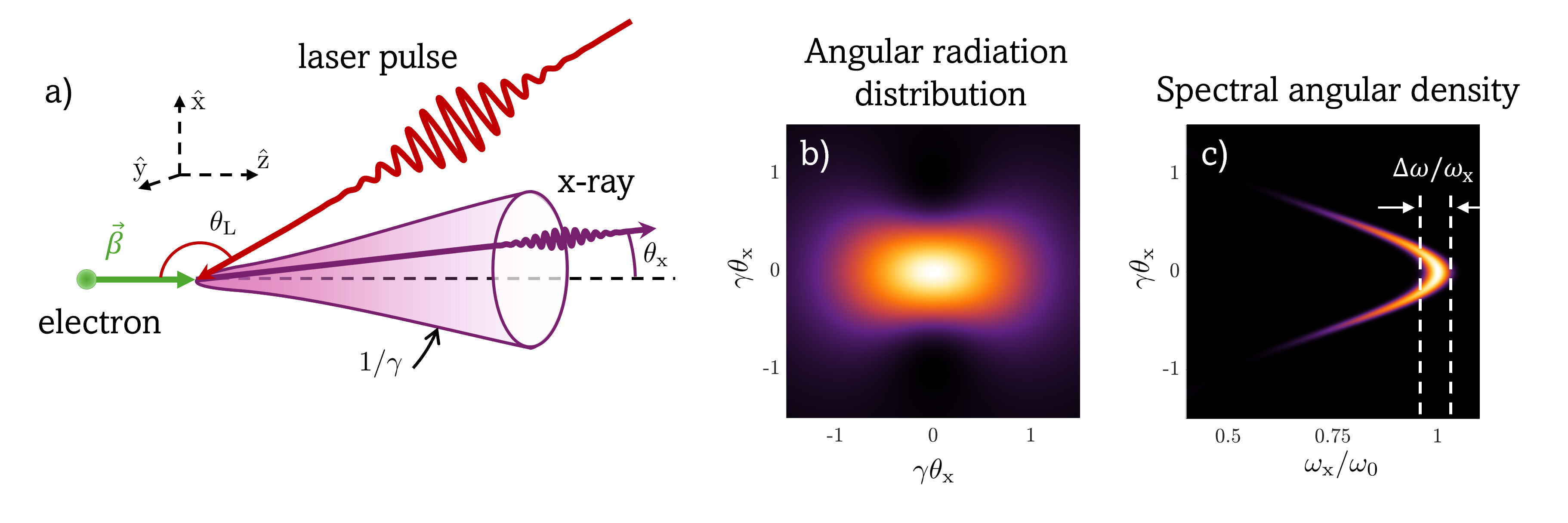}
    \caption{a) Schematic representation of ICS. An electron (green) interacts with a laser pulse (red) under an angle $\theta_L$, generating x-rays with emission angle $\theta_\mathrm{x}$. b) shows the angular radiation distribution, which becomes cylindrically symmetric for $\theta_\mathrm{x} \ll 1/\gamma$. Here the vertical axis is in the direction of the laser polarization. c) shows the spectral angular density, with $\omega_0 = \omega_L (1-\beta \cos\theta_L)/(1-\beta)$ the x-ray frequency for $\theta_\mathrm{x}=0$}
    \label{fig:ICS_geometry}
\end{figure*}
\label{sec:introduction}
Coherent and tunable x-ray sources have become an indispensable tool in many fields of research, such as materials science \cite{sedigh2020review}, life sciences \cite{thomlinson2018k}, and advanced medical diagnostics \cite{quenot2022x}. This demand is primarily served by large facilities such as synchrotrons and free electron lasers \cite{couprie2014new}. However, due to the limited availability of these large facilities, their usage comes with many logistical challenges. Therefore, there is a growing need for coherent and tunable x-ray sources at the scale of the university lab.\\ 

Inverse Compton scattering (ICS), a phenomenon in which a highly energetic electron collides with a photon in such a way that the electron looses energy and the photon gains energy, forms the basis of a promising method to generate x-rays at the lab scale. Several ICS based x-ray sources are currently operational \cite{gunther2020versatile,du2013generation,alkadi2025commissioning,graves2014compactburstmode,VanElk:25,barty2024design,amoudry2025commissioning} and new designs have been proposed \cite{faillace2019status,brummer2022compact,musat2024a,schaap2024inverse,deitrick2018high}. Imaging techniques that require energy tunability and coherence, such as K-edge subtraction imaging \cite{gunther2025rapid} and phase contrast imaging \cite{Berthe_2024_PhaseContrast}, have been demonstrated with an ICS based x-ray source and the use of lensless techniques such as ptychography has been proposed \cite{batey2021x}. \\

While the fundamental theory of ICS has been described in great detail \cite{hartemann2013nonlinear, esarey1993nonlinear, schaap2022photon,krafft2010compton}, a closed form analytical treatment, that takes into account both the three dimensional distribution of the pulsed electron and laser beams and arbitrary interaction angles, is lacking. Optimization tasks are therefore left to simulations, which can be time consuming and do not provide a thorough physical understanding of the limitations.\\

The goal of this paper is to provide an analytical closed form theoretical framework to optimize x-ray generation by ICS.  We restrict ourselves to the Thomson regime, where electron recoil can be ignored.  Different applications require optimization of different x-ray beam properties. In this paper we choose to optimize the x-ray brilliance, which is a comprehensive figure of merit that take into account x-ray flux, transverse coherence and energy spread. The framework can be extended, however, to optimize other quantities such as flux of spectral density. This is described in Appendix \ref{app:quantities}. The average x-ray brilliance is defined as \cite{attwood2000soft}
\begin{equation}\label{eq:Brilliance}
    B_\mathrm{x} = \frac{f N_\mathrm{x}}{\Delta A\Delta\Omega (\Delta\omega/\omega_\mathrm{x})}.
\end{equation}
Here, $N_\mathrm{x}$ is the number of x-rays per pulse, $f$ is the pulse repetition frequency, $\Delta A$ is the effective x-ray emission area, and $\Delta \Omega$ is the solid angle in which the x-rays are emitted. $\Delta\omega/\omega_\mathrm{x}$ is the relative bandwidth of the x-ray beam. For details on the definition of the x-ray brilliance used in the paper, see Appendix \ref{app:Brilliance}.\\

The framework is meant to provide an intuitive understanding of how the x-ray source performance scales with the parameters of the interaction geometry. Here, the term \textit{geometry} refers not only to the interaction angle, but also to the electron and laser pulse dimensions, the electron beam energy, and the bunch charge. The framework could serve as a baseline for a Compton x-ray source design, in which, given a desired x-ray energy and performance of the electron beamline and laser system, the optimized interaction geometry can be calculated. \\

%%%%%%%%%%%%%%%%%%%%%%%%%%%%%%%%%%%%%%%%%%%%%%%%
%%%%%%%%%%%%%%%%%%%%%%%%%%%%%%%%%%%%%%%%%%%%%%%%
\subsection{Basics of inverse Compton scattering}

Figure \ref{fig:ICS_geometry}a shows a schematic illustration of the ICS geometry. An electron (green) with velocity $\vec{v}= \beta c \rm\hat{z}$ and normalized energy $\gamma = 1/\sqrt{1-\beta^2}$ interacts with a laser pulse (red) under an interaction angle $\theta_L$, w.r.t. the electron velocity. This angle is defined in such a way that $\theta_L = \pi$ corresponds to a head-on collision. During the interaction, the electron oscillates rapidly and radiates in the process (purple). Due to the relativistic nature of the electron, the radiation becomes strongly blue shifted. In the Thomson regime, the shifted frequency is given by \cite{schaap2022photon} 
\begin{align}\label{eq:dopplerShift1}
    \omega_\mathrm{x} = \omega_L \frac{1-\beta \cos\theta_L}{1- \beta \cos\theta_\mathrm{x}}.
\end{align}
Here, $\omega_\mathrm{x}$ and $\omega_L$ are the angular frequencies of the generated radiation and the laser pulse, respectively, and $\theta_\mathrm{x}$ is the angle of the outgoing radiation with respect to the electron velocity. \\

In the relativistic limit ($\gamma \gg1$) and under small radiation angles ($\gamma \theta_\mathrm{x} \ll 1$), Eq.~\eqref{eq:dopplerShift1} can be approximated in a convenient way.  For head-on scattering $(\theta_L = \pi)$ this results in the well known expression
\begin{equation}\label{Eq:DopplerHO}
   \omega_{\mathrm{x}} \underset{\theta_L= \pi}{\simeq} \omega_L\frac{4\gamma^2 }{1+\gamma^2\theta_\mathrm{x}^2},
\end{equation}
whereas for small interaction angles $(\theta_L \ll 1)$ we find
\begin{equation}\label{Eq:DopplerGrazing}
   \omega_{\mathrm{x}} \underset{\theta_L\ll 1}{\simeq} \omega_L\frac{1+\gamma^2\theta_L^2}{1+\gamma^2\theta_\mathrm{x}^2}.
\end{equation}
Note that, since typically a large frequency shift is desired, preferably $\gamma\theta_\mathrm{L} \gg 1 $. These approximations clearly show that the x-ray energy can be tuned by varying either the electron energy, $\gamma$, or the interaction angle $\theta_L$.\\

Furthermore, \cref{Eq:DopplerHO,Eq:DopplerGrazing} show how the x-ray frequency depends on the angle $\theta_\mathrm{x}$: the highest x-ray frequency, $\omega_0$, is emitted along the propagation direction of the electron ($\theta_\mathrm{x} = 0$) and the x-ray frequency gradually decreases for larger angles, as shown in Fig~\ref{fig:ICS_geometry}c \cite{sakai2017single}. At $\theta_\mathrm{x} \simeq 1/\gamma$ the x-ray frequency has become half of the on-axis frequency, $\omega_0$. Fig~\ref{fig:ICS_geometry}b shows the intensity distribution of ICS, which is concentrated in a narrow cone characterized by a half angle $ \theta_\mathrm{x} = 1/\gamma$ \cite{albert2010characterization,jackson_classical_1999}. The distribution is slightly asymmetric for linear laser polarization, which is pointing in the vertical direction in Fig~\ref{fig:ICS_geometry}b \cite{VanElk:25,taira2025generation}. However, for $ \theta_\mathrm{x} \ll 1/\gamma$, which is required for a narrow bandwidth x-ray beam, the angular distribution of the radiation is to a good approximation uniform.\\

The central challenge in the development of Compton x-ray sources lies in the low efficiency of x-ray generation per electron, associated with the small Thomson cross section, 
\begin{equation}
    \sigma_T = \frac{8 \pi}{3} r_e^2 \simeq 66.5 \ \mathrm{fm^2},
\end{equation}
where $r_e = e^2/(4\pi\epsilon_0 mc^2)$ is the classical electron radius, with $e$ the elementary charge and $m$ the electron mass. Obviously, for efficient generation of x-rays by Compton scattering, the density of electrons and laser photons needs to be as high as possible, requiring focused electron and laser pulses.\\

%%%%%%%%%%%%%%%%%%%%%%%%%%%%%%%%%%%%%%%%%%%%%%%%
%%%%%%%%%%%%%%%%%%%%%%%%%%%%%%%%%%%%%%%%%%%%%%%%
\subsection{Structure of this paper}
We will analyze the dependence of the x-ray generation on geometry of ICS in increasing levels of complexity, illustrated by Fig.~\ref{fig:structureofpaper}. We start in Sec.~\ref{sec:head-on ics} by describing ICS in a head-on geometry using a collisional model. We first consider the collision between pencil beams, i.e. in the approximation of perfectly parallel beams (\Circled{1} in Fig.~\ref{fig:structureofpaper}). Next, we analyze the more realistic situation of a focused laser pulse and a pencil electron beam (\Circled{2} in Fig.~\ref{fig:structureofpaper}) and find the optimized interaction spot size and the corresponding optimized bunch charge.\\

In Sec.~\ref{sec:GrazingAngleICS}, we analyze ICS for arbitrary interaction angles. This is the most novel part of the paper. A proper analysis of this geometry requires a treatment in a covariant framework. Again, we first analyze the situation of two pencil beams (\Circled{3} in Fig.~\ref{fig:structureofpaper}). Subsequently, we investigate the combination of a focused laser pulse and a pencil electron beam (\Circled{4} in Fig.~\ref{fig:structureofpaper}). We derive closed form analytical expressions for the laser focal spot size and the electron bunch charge, which optimize the x-ray brilliance in a grazing angle geometry. We find that a shallow co-propagating interaction angle -- a \textit{grazing angle geometry} -- can be favorable compared to a head-on geometry, especially for the generation of soft x-rays.\\ 

\begin{figure}[h]
    \centering
     \includegraphics[width=\linewidth]{Geometries.png}
    \caption{Schematic illustrations of the four geometries considered in this paper. In each geometry the electron bunch (green) with velocity $\boldsymbol\beta$ and laser pulse (red) with wave vector $\boldsymbol k_L$ are shown at three different times up until the collision. The top figures show the pencil beam \Circled{1} and focused laser beam \Circled{2} geometries in a head-on geometry. The bottom figures show the pencil beam \Circled{3} and focused laser beam \Circled{4} geometries in a grazing angle geometry. The relevant r.m.s. pulse dimensions are also indicated }
    \label{fig:structureofpaper}
\end{figure}

%%%%%%%%%%%%%%%%%%%%%%%%%%%%%%%%%%%%%%%%%%%%%%%%
%%%%%%%%%%%%%%%%%%%%%%%%%%%%%%%%%%%%%%%%%%%%%%%%
\section{Head-on inverse Compton scattering}
\label{sec:head-on ics}
In this section we describe ICS in a head-on geometry $(\theta_L = \pi)$. We start with a pencil beam description and calculate the x-ray brilliance. The result from this simplified model serves as a demonstration of important scalings for ICS. Next, we consider a finite laser pulse length to find the optimal laser waist. Finally, a simple argument is used to determine the bunch charge that maximizes the x-ray brilliance.

\subsection{Pencil beam description}\label{sec:headon_pencil}
We assume the pulses to be pencil beams with a Gaussian transverse photon/electron density distribution
\begin{equation}\label{eq:DensityIdeal}
    \rho_{L,e} = \frac{N_{L,e}}{2\pi \sigma^2_{L/e}}\exp\left(-\frac{x^2+y^2}{2\sigma_{L/e}^2}\right),
\end{equation}
where the subscripts $L$ and $e$ refer to the laser beam and the electron beam, respectively. $N_{L/e}$ is the total number of laser photons/electrons and $\sigma_{L/e}$ is the r.m.s. transverse size. Throughout this paper, a quantity $\sigma$ refers to an r.m.s. quantity (with the exception of the Thomson cross section $\sigma_T$). Note that we can neglect the $z$-dependence of the density distributions due to the head-on geometry and the pencil beam approximation. The total number of scattered x-rays in the collision is given by \cite{furman2003moller}
\begin{equation}\label{eq:xrayTot1}
    N_\mathrm{x} = \sigma_\mathrm{T}\int dt \int d \mathbf{x} \rho_L \rho_\mathrm{e}|v_\mathrm{e}-v_L| = \frac{ \sigma_\mathrm{T}N_L N_\mathrm{e}}{2 \pi (\sigma^2_\mathrm{e}+\sigma^2_L)},
\end{equation}
Note that Eq.~\eqref{eq:xrayTot1} corresponds to the situation in which the x-rays are scattered in the complete $4\pi$ solid angle with 100\% relative energy spread. If a small relative energy spread $\Delta\omega/\omega_\mathrm{x}$ is required, the x-ray acceptance angle should be reduced to $\Theta_\mathrm{x} = \sqrt{(\Delta\omega/\omega_\mathrm{x})} /\gamma$. If $\gamma\Theta_\mathrm{x} \ll1$, the number of x-rays in this acceptance angle is \cite{jackson_classical_1999,krafft2010compton}
\begin{equation}\label{eq:xrayTotBW}
    N_\mathrm{x}  = \frac{\sigma_\mathrm{T}N_L N_\mathrm{e}}{4 \pi \sigma^2_\mathrm{e}}\frac{3}{2} \gamma^2\Theta_\mathrm{x}^2.
\end{equation}
Here, we have additionally assumed that the sizes of the electron and laser pulses are matched ($\sigma_L = \sigma_\mathrm{e}$), which is close to the optimized situation and simplifies the analysis. This is an important assumption used throughout this section. Using $\Delta A \Delta\Omega = 2\pi^2 \sigma_\mathrm{e}^2\Theta_\mathrm{x}^2 $ and $\Delta\omega/\omega_\mathrm{x} = \sqrt{2\pi} \sigma_\omega/\omega_\mathrm{x}$, with $ \sigma_\omega/\omega_\mathrm{x}$ the r.m.s. relative bandwidth, the average brilliance is given by 
\begin{equation}\label{eq:BrillianceTot}
        %B_\mathrm{x}  = \frac{ f\sigma_\mathrm{T}N_L N_\mathrm{e}}{8 \pi^3 } \frac{3}{2}\frac{1}{\sigma_e^4\Theta_\mathrm{x}^2}\frac{\gamma^2\Theta_\mathrm{x}^2}{\Delta\omega/\omega_\mathrm{x}}.
        B_\mathrm{x}  = \frac{ f\sigma_\mathrm{T}N_L N_\mathrm{e}}{\sqrt{2\pi }\ 8 \pi^3 } \frac{3}{2}\frac{1}{\sigma_e^4\Theta_\mathrm{x}^2}\frac{\gamma^2\Theta_\mathrm{x}^2}{\sigma_\omega/\omega_\mathrm{x}}.
        % B_\mathrm{x}  = \frac{ f\sigma_\mathrm{T}N_L N_\mathrm{e}}{8 \pi^3 \sigma_e^4 \Theta_\mathrm{x}^2 } \frac{3}{2}\frac{\gamma^2\Theta_\mathrm{x}^2}{\Delta\omega/\omega_\mathrm{x}}.
\end{equation}

The energy spread of a Compton x-ray source has several contributions originating from both the laser pulse and the electron beam properties \cite{petrillo2012photon}. To illustrate important scalings, we restrict ourselves in the first instance to the contribution to the energy spread due to the angular spread of the electron beam: $\left.\sigma_{\omega}/\omega_\mathrm{x}\right|_{\sigma_{\theta e}} = \gamma^2\sigma_{\theta e}^2 $. If we describe the electron beam using the normalized emittance, $\epsilon_n = \gamma\beta \sigma_e\sigma_{\theta e} \simeq \gamma \sigma_e\sigma_{\theta e}$ , Eq.~\eqref{eq:BrillianceTot} can be rewritten as
\begin{equation}\label{eq:BrillianceTot2}
        B_\mathrm{x}  = \frac{ f\sigma_\mathrm{T}N_L N_\mathrm{e}}{\sqrt{2\pi}\ 8 \pi^3 } \frac{3}{2}\frac{\gamma^4 \sigma_{\theta e}^2}{\epsilon_n^4}.
\end{equation}
Equation \eqref{eq:BrillianceTot2} shows two important scalings: First, we find that the brilliance scales with $\gamma^4$, which originates from the improved geometric emittance with increased electron energy. This not only results in a smaller x-ray source size, but also allows for tighter focusing of the laser pulse. In a head-on geometry, the electron energy is fixed by the desired x-ray energy for a given laser wavelength. However, if the interaction angle can be varied, this scaling can be exploited for any desired x-ray energy. This will be further explored in Sec.~\ref{sec:GrazingAngleICS}.\\

Second, Eq.~\eqref{eq:BrillianceTot2} shows that $B_\mathrm{x} \propto N_\mathrm{e} / \epsilon_n^4$. For space charge dominated electron beams, it holds that $N_\mathrm{e}/\epsilon_{\mathrm{n}}^2 = \mathrm{constant.}$ As a result we find that the x-ray brilliance increases with a lower bunch charge. This counterintuitive result is a direct consequence of the assumption that the spot size of the electron and laser beams are matched $(\sigma_L = \sigma_e)$, which is an assumption that cannot hold indefinitely, as it is obviously limited by the laser beam diffraction limit. In the remainder of this section we calculate the x-ray flux and brilliance with realistic electron and laser beam descriptions, and find the optimized laser focus and the corresponding bunch charge to match the electron spot size to the laser pulse.\\

%%%%%%%%%%%%%%%%%%%%%%%%%%%%%%%%%%%%%%%%%%%%%%%%
%%%%%%%%%%%%%%%%%%%%%%%%%%%%%%%%%%%%%%%%%%%%%%%%
\subsection{Focused laser pulse}\label{sec:divergence}
\begin{figure}
    \centering
     \includegraphics[width=\linewidth]{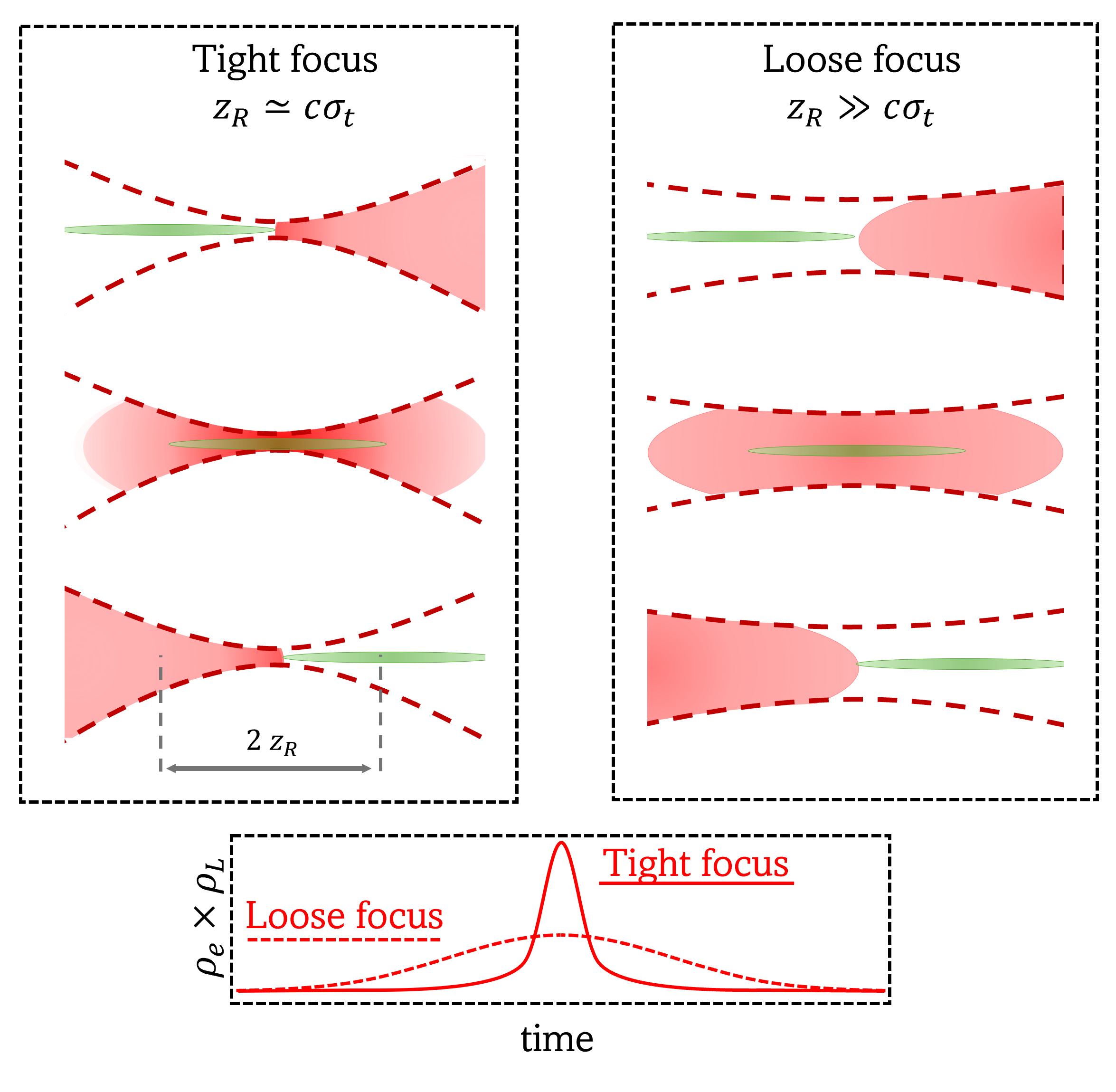}
    \caption{Illustration showing how focusing the laser beam results in a reduction of the Rayleigh length, which affects the interaction time. The bottom graph shows the product of the electron pulse density and laser pulse photon density as a function of time for a tight focus and a loose focus.}
    \label{fig:laserDivergence}
\end{figure}

Before the laser beam diffraction limit is reached, the Rayleigh range may become comparable to, or even smaller than, the laser pulse length, resulting in a significant reduction of the interaction time between the electron pulse and the laser pulse (see Fig. \ref{fig:laserDivergence}). This will lead to both a reduction of the number of x-ray photons produced, and an increase of the x-ray energy spread. To analyze the effect of focusing the laser beam we now consider the photon density of a $\mathrm{TEM}_{00}$ pulsed Gaussian laser beam
\begin{equation}\label{eq:densityDivergence}
   \rho_L = \frac{N_L}{(2\pi)^{3/2}  \sigma^2_L(z)\ \sigma_t}\exp\left[-\frac{x^2+y^2}{2\sigma_L^2(z)}  -\frac{(z/c + t)^2}{2 \sigma_t^2}\right],
\end{equation}
where $\sigma_t$ is the laser pulse length and $\sigma_L(z) = \sigma_L(0)\sqrt{1+(z/z_R)^2}$ with $z_R = 2 k_L \sigma_L^2$ the Rayleigh length. Throughout this paper we assume perfect laser beams ($M^2 =1$). For non-perfect laser beams the analysis still holds if the expressions for the Rayleigh length and the angular spread are adjusted accordingly. For now, we assume the electron beam to be described by a pencil beam. Substituting \eqref{eq:densityDivergence} into the central term of Eq.~\eqref{eq:xrayTot1} and subsequently performing the integration over time results in

\begin{equation}\label{eq:xrayTotDiv}
    N_\mathrm{x} =  \frac{ \sigma_\mathrm{T}N_L N_\mathrm{e}}{4 \pi  \sigma^2_\mathrm{e}} \frac{3}{2} \gamma^2 \Theta_\mathrm{x}^2\sqrt{\pi}\  u \ \mathrm{erfcx}(u).
\end{equation}
Here, $\mathrm{erfcx}(x) = \exp(x^2)\mathrm{erfc}(x) $ is the scaled complementary error function \cite{abramowitz1948handbook}, and $u = 2z_R/c\sigma_t$. The term  $\sqrt{\pi}\  u \ \mathrm{erfcx}(u)$ describes how the effective interaction time is reduced due to laser beam divergence \cite{hartemann2013nonlinear}. If we assume $\epsilon_n /\gamma \ll \lambda_L/4\pi$, a very reasonable assumption in practice, then also the divergence and finite longitudinal size $\sigma_z$ of the electron beam can be taken into account using $u = 2 z_R/\sqrt{(c^2 \sigma_t^2 + \sigma_z^2) }$.\\

To calculate the brilliance, we must now take into account the x-ray energy spread, to which both the laser beam properties and the electron beam properties contribute. Here we consider only the three contributions of the laser beam. First, we take into account the Fourier limited bandwidth $ \left.\sigma_{\omega}/\omega_\mathrm{x}\right|_{\omega_L} = 1/(2\omega_L\sigma_t) $, which  is broadened due to the shorter interaction time: 
\begin{equation}\label{eq:BW_laser1}
   \left.\sigma_{\omega}/\omega_\mathrm{x}\right|_{\omega_L} = 1/\left[ 2\omega_L \sigma_t \sqrt{\pi}\  u \ \mathrm{erfcx}(u) \right].
\end{equation}

Second, the angular spread of a focused laser pulse gives rise to a spread in the interaction angle $\sigma_{\theta L} = 1/(2 k_L\sigma_L)$, resulting in additional x-ray broadening. By expanding Eq.~\eqref{eq:dopplerShift1} around $\theta_L = \pi$ we find $ \left.\sigma_{\omega}/\omega_\mathrm{x}\right|_{\sigma_{\theta L}} = 1/\left(2 k_L \sigma_L\right)^2 $ \cite{petrillo2012photon}, which can be expressed in terms of $u$ as
\begin{equation}\label{eq:BW_laser_div}
   \left.\sigma_{\omega}/\omega_\mathrm{x}\right|_{\sigma_{\theta L}} ={1 \over \omega_L \sigma_t u}
\end{equation}

Finally, at high field intensities the x-ray energy spread is increased due to nonlinear broadening \cite{hartemann2010low}, characterized by the normalized vector potential amplitude $A_0 = eE_0/(mc\omega_L)$. The corresponding x-ray energy spread is given by \cite{petrillo2012photon}
\begin{equation}\label{eq:BW_A0}
   \left.\sigma_{\omega}/\omega_\mathrm{x}\right|_{A_0} \simeq \frac{1}{6}A_0^2 ={\chi \over \omega_L \sigma_t u}.
\end{equation}
In the right hand side term we have expressed the nonlinear broadening in terms of the laser pulse dimensions. Here, $\chi = 4 \sigma_T N_L/(3\sqrt\pi \alpha \lambda_L c \sigma_t)$ relates the laser peak power to the normalized vector potential and $\alpha$ is the fine structure constant. $\chi$ is only dependent on the laser pulse energy and duration, and \textit{not} the focal spot size. The nonlinear broadening can become the significant contribution to the x-ray energy spread if the laser is tightly focused. We note that the nonlinear broadening can be circumvented using appropriate laser pulse chirping \cite{terzic2021laser,terzic2020improving,krafft2023scattered}, which would effectively bring $\chi \rightarrow0$.  \\

The three energy spread contributions associated with the laser beam properties can be combined, resulting in the following expression for the x-ray brilliance
\begin{equation}
    \label{eq:BrillianceDivergence1}
     B_\mathrm{x}  = \frac{ 3 f\sigma_\mathrm{T}N_L N_\mathrm{e}}{\sqrt{2\pi}\ 8 \pi^3 } \frac{\gamma^2  \omega_L \sigma_t}{\sigma_e^4} \frac{\pi u^2 \ \mathrm{erfcx}(u)^2}{\sqrt{1+ 4\pi(1+\chi^2)\mathrm{erfcx}(u)^2 }} .
\end{equation}

To optimize brilliance, we note that for space charge limited electron beams $\epsilon_{n}  = \eta \sqrt{e N_e}$ \cite{pasmans2016extreme}, where $\eta$ is a property of the injector that describes the electron beam emittance at given bunch charge. Therefore $N_e/\sigma_e^2 = \mathrm{constant.}$ Since  $u \propto \sigma_L^2 =\sigma_e^2$, the function  $g(u) =   \pi u\ \mathrm{erfcx}(u)^2/\sqrt{1+ 4\pi(1+\chi^2)\mathrm{erfcx}(u)^2 }$ must be maximized, resulting in the optimized value for $\sigma_e$
\begin{equation}\label{eq:sigeopt}
    \sigma_e^{\mathrm{opt}} = \sqrt{u_\mathrm{max} \lambda_L c \sigma_t  /8 \pi\  }
\end{equation}
where $u_\mathrm{max} $ is the value of $u$ that maximizes $g(u)$ and thus the x-ray brilliance. To take into account the length of the electron pulse $\sigma_z$, the substitution $\sigma_t \rightarrow \sigma_t \sqrt{1+ (\sigma_z/c\sigma_t)^2 }$ should be used in Eq.~\eqref{eq:sigeopt}.  \\

\cref{eq:BrillianceDivergence1,eq:sigeopt} do not take into account bandwidth contributions originating from the electron energy spread, $ \left.\sigma_{\omega}/\omega_\mathrm{x}\right|_{\sigma\gamma} = 2\sigma_\gamma/\gamma$, where $\sigma_\gamma$ is the electron energy spread, and the electron angular spread, $ \left.\sigma_{\omega}/\omega_\mathrm{x}\right|_{\sigma_\theta e} = \gamma^2\sigma_{\theta e}^2$. To include these contributions, the substitution $\chi^2 \rightarrow \chi^2 + \nu^2 $ should be made, where
\begin{equation}
    \nu =\omega_L\sigma_t u \sqrt{(2\sigma_\gamma/\gamma)^2 + (\gamma^2\sigma_{\theta e}^2)^2} .
\end{equation}
 This has the consequence of decreasing the value of $u_\mathrm{max}$\\
 
\begin{figure}
    \centering
     \includegraphics[width=\linewidth]{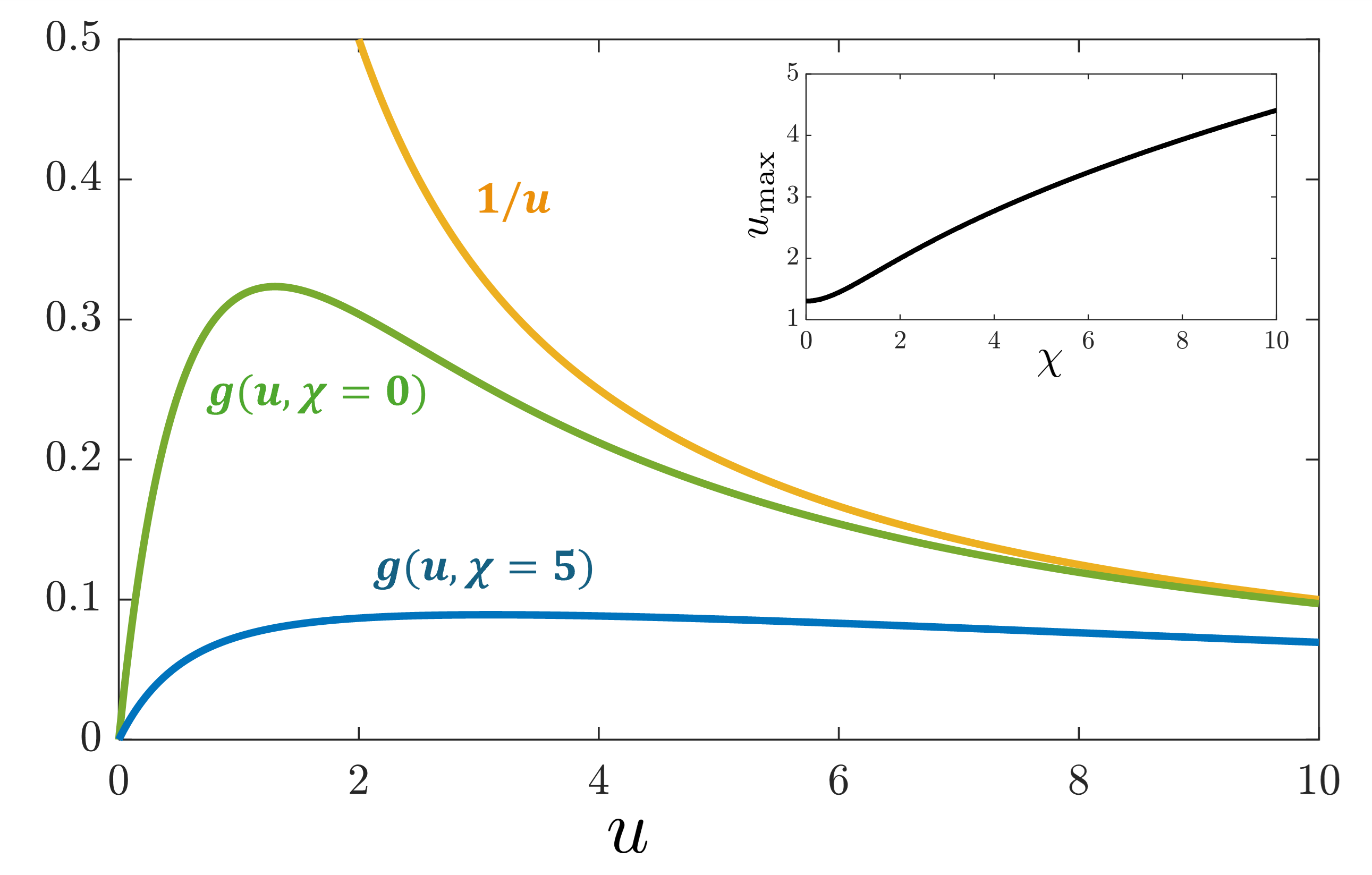}
    \caption{Plots of the function $g(u)$ for $\chi =0$ (green curve) and $\chi = 5$ (blue curve). The inset shows a plot of $u_\mathrm{max}$ as a function of $\chi$. For comparison the function $1/u$ (yellow curve) is plotted as well, which gives the brilliance scaling of the pencil beam limit.}
    \label{fig:erfcx_plot}
\end{figure}

In Fig.~\ref{fig:erfcx_plot} the function $g(u)$ is plotted for $\chi =0$ (green curve) and $\chi = 5$ (blue curve), respectively. The function $1/u$ (yellow curve) is plotted as well, which gives the brilliance scaling in the pencil beam limit. The inset shows a plot of $u_\mathrm{max}$ as a function of $\chi$.\\

This analysis has been performed under the assumption that the electron and laser spot sizes are matched, i.e., $\sigma_e =\sigma_L$. For a fixed energy, the electron spot size is determined by the beam divergence $\sigma_{\theta e}$ and the normalized emittance $\epsilon_n$. For relativistic beams $\sigma_{\theta e}$ is typically limited in practice to a few milliradians. We choose a fixed electron beam divergence of 1 mrad and optimize the normalized emittance such that $\sigma_e =\sigma_L$. To arrive at an expression for the optimized bunch charge we use the fact that $\epsilon_{n}  = \eta \sqrt{e N_e} = \gamma \sigma_e \sigma_{\theta e}$, combined with Eq.~\eqref{eq:sigeopt}. This results in
\begin{equation}\label{eq:Qopt}
    e N_{e}^{\mathrm{opt}} = u_\mathrm{max}\gamma^2 \sigma^2_{\theta e} \frac{ \lambda_L c \sigma_t}{8 \pi  \eta^2},
\end{equation}
which is the bunch charge that allows for the optimized electron spot size $\sigma_e^{\mathrm{opt}}$.\\

Interestingly, by substitution of \cref{eq:sigeopt,eq:Qopt} into Eq.~\eqref{eq:BrillianceTot2} we find that the x-ray brilliance is independent on the pulse lengths. However, This is only the case when the energy spread contributions of \textit{only} the laser pulse are included. When including the energy spread contributions of the electron beam, the laser pulse length is optimized when the total of the energy spread contributions of the laser pulse is equal to the energy spread contributed by the electron beam.\\

We note that the presented optimization assumes ideal Gaussian electron and laser beams. To account for non-ideal laser beams, the substitution $z_R\rightarrow z_R/M^2$ should made in the definition of $u=z_R/2c\sigma_t$. Similarly, non-ideal electron beams can be accounted for by increasing normalized emittance to account for the increased interaction spot size. When the laser pulse has time bandwidth product above the Fourier limit, Eq.~\eqref{eq:BW_laser1} should be scaled accordingly. Note that the definition of $u$ remains unchanged, since the pulse length remains the same.\\

%%%%%%%%%%%%%%%%%%%%%%%%%%%%%%%%%%%%%%%%%%%%%%%%
%%%%%%%%%%%%%%%%%%%%%%%%%%%%%%%%%%%%%%%%%%%%%%%%
\subsection{Optimized head-on geometry}
As an example, we calculate the brilliance of an ICS source generating 10 keV x-rays in a head-on geometry. The electron bunches are generated from a C-band photoinjector \cite{lucas2023toward,alesini2026design} that is able to create 200 pC electron bunches at $\epsilon_n = 200 \mathrm{\ nmrad}  \ (\eta = 14\mathrm{\ nmrad/\sqrt{pC}}) $ and a relative energy spread of 0.2\%. The electrons are accelerated to $\gamma = 46$ to generate 10 keV x-rays with a 5 mJ, 100 fs, 1030 nm interaction laser pulse ($\chi = 5.6$) \cite{LightConversion_Pharos}. In this example, we take the electron divergence angle to be $\sigma_\theta = 1$ mrad, and the electron pulse length and the laser pulse length are equal. Both the photoinjector and the laser operate at $f = 1$~kHz. \\

\begin{figure}
    \centering
     \includegraphics[width=\linewidth]{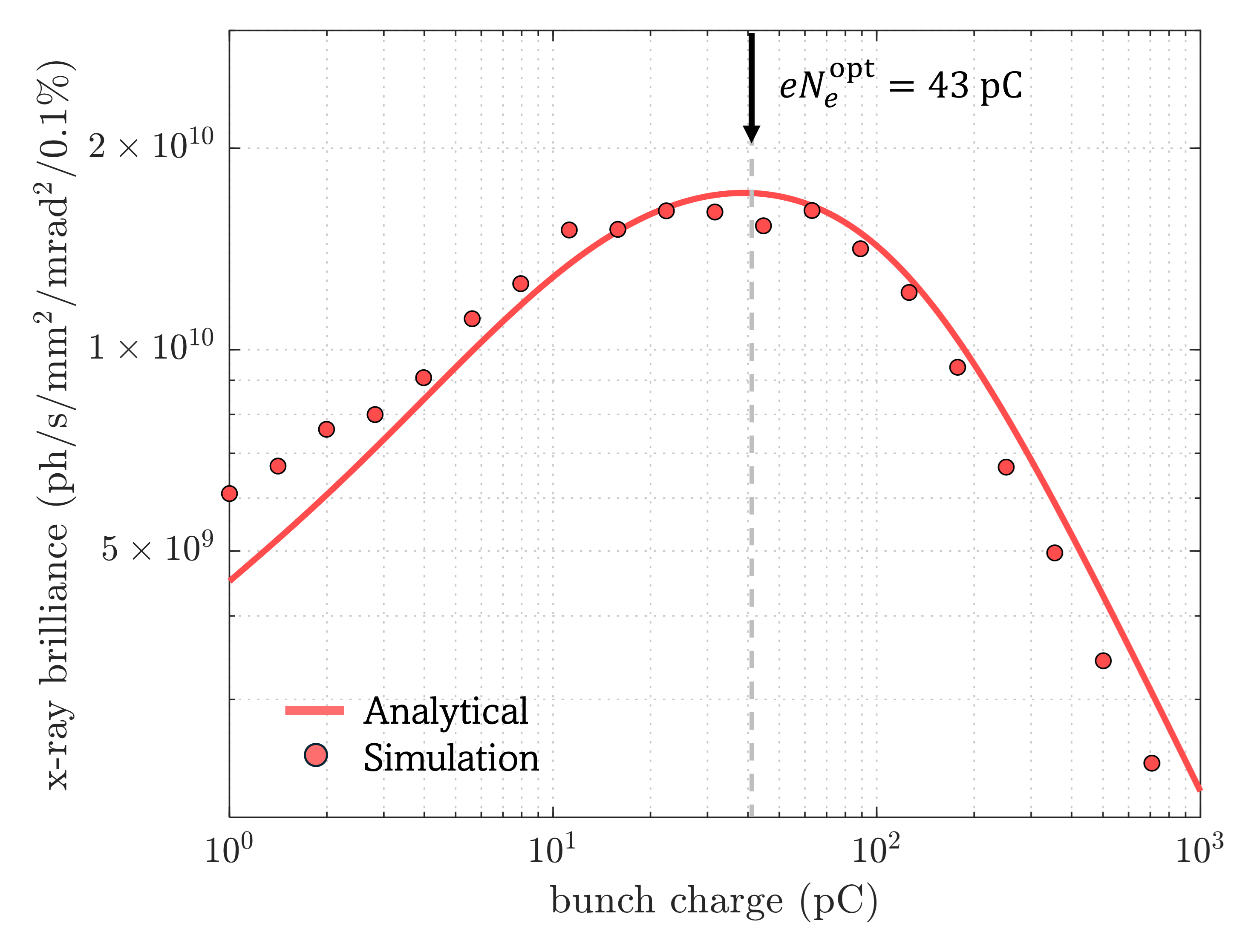}
    \caption{X-ray brilliance for 10 keV x-ray generation in a head-on geometry with matched electron and laser spot sizes, as a function of bunch charge. The analytical result of Eq.~\eqref{eq:BrillianceDivergence1} (solid curve) agrees well with simulation results (dots). The arrow marks the optimized bunch charge given by Eq.~\eqref{eq:Qopt} }
    \label{fig:varyQ_plot}
\end{figure}

Figure \ref{fig:varyQ_plot} shows the x-ray brilliance as a function of bunch charge. Here the bunch charge determines $\sigma_e = \eta \sqrt{eN_e}/(\gamma \sigma_{\theta e})$ and the size of the laser is matched to the electron beam, $\sigma_L = \sigma_e$. This results in a bunch charge where $\sigma_L $ and $\sigma_e$ are optimized. The energy spread contributions due to the electron beam properties have also been included. The continuous curve is calculated using Eq.~\eqref{eq:BrillianceDivergence1}, and the dots are the result of particle tracking simulations (General Particle Tracer \cite{PulsarPhysicsGeneralGPT}) and subsequent numerical calculation of the Liénard-Wiechert potentials \cite{thomas2010algorithm}. From \cref{eq:sigeopt,eq:Qopt} it then follows that $\sigma_e^{\mathrm{opt}} = 2.4\  \mathrm{\mu m}$ and the corresponding optimized bunch charge $ e N_{e}^{\mathrm{opt}} =43\,{\rm pC}$.\\

At the relatively low bunch charge of 43 pC, the x-ray brilliance is 10 times higher than at 1 nC. Additionally, the total number of scattered x-rays is only  4\%  lower at 43 pC than at 1 nC. Remarkably, due to the optimization of the laser focus and bunch charge, ICS is significantly more efficient at a bunch charge which is $\sim$20 times lower. \\

%%%%%%%%%%%%%%%%%%%%%%%%%%%%%%%%%%%%%%%%%%%%%%%%
%%%%%%%%%%%%%%%%%%%%%%%%%%%%%%%%%%%%%%%%%%%%%%%%
\section{inverse Compton scattering for arbitrary interaction angles}\label{sec:GrazingAngleICS}
In a head-on ICS geometry, the generated x-ray energy increases with the electron energy. For a given laser wavelength, the desired x-ray energy therefore determines the electron energy. At the same time, the x-ray brilliance increases with electron energy, implying that, using ICS, the generation of hard x-rays is more favorable than the generation of soft x-rays. However, if instead of the electron energy, the interaction angle is varied to achieve the required x-ray energy, an additional degree of freedom is introduced, which allows further exploitation of the favorable scaling of the brilliance with the electron beam energy. This will be explored in some detail in this section. We will find that a grazing angle geometry (see Fig.~\ref{fig:ICS_geometry}) can be particularly useful for the generation of high-brilliance soft x-rays. \\

We will now use the framework of covariant electrodynamics, with the metric $g^{\mu \nu} = \mathrm{diag}(1,-1,-1,-1)$, to describe the interaction of a relativistic electron bunch with a laser pulse under an arbitrary interaction angle. We start, in Sec.~\ref{sec:colin_singleelec}, by considering a single electron interacting with a Gaussian pencil beam. Next, in Sec.~\ref{sec:colin_divergence}, we investigate the effects of focusing the laser beam, and in Sec.~\ref{sec:Focused electron and laser beams}, an electron bunch of finite dimensions instead of a single electron. In Sec.~\ref{sec:optimization} the x-ray brilliance is optimized for $\theta_L \ll1 $. Finally, in Sec.~\ref{sec:performance}, the optimized performances of ICS in a grazing angle geometry and in a head-on geometry are compared.\\

%%%%%%%%%%%%%%%%%%%%%%%%%%%%%%%%%%%%%%%%%%%%%%%%
%%%%%%%%%%%%%%%%%%%%%%%%%%%%%%%%%%%%%%%%%%%%%%%%
\subsection{Single electron with a pencil laser beam}\label{sec:colin_singleelec}
In the regime where electron recoil can be ignored, \textit{i.e.}, $4\gamma^2(1-\cos\theta_L)\hbar\omega_L/\gamma mc^2\ll 1$ \cite{serafini2024compton}, a single electron with 4-trajectory $x^\mu(\tau)$ and 4-velocity $u^\mu(\tau)$ radiates, according to the Liénard-Wiechert potentials, with a spectral angular density \cite{hartemann2013nonlinear}
\begin{equation}
    \label{eq:specangCovariant}
    \mathcal{S}(\omega, \theta_\mathrm{x}) = \frac{\partial^2N_\mathrm{x}}{\partial\omega\partial\Omega}= \frac{\alpha \omega_\mathrm{x}}{4 \pi^2 } \left|\int u^\mu \exp{(i k_\mathrm{x}^\nu x_\nu)} d\tau  \right|^2
\end{equation}
Here $k_\mathrm{x}^\mu = k_\mathrm{x}(1,\boldsymbol{n}_\mathrm{x})$ is the four-wave vector of the scattered x-rays with $\boldsymbol{n}_\mathrm{x}$ the propagation direction. Furthermore, $x^\mu = (c t, \boldsymbol{x} )$ is the electron four-position, $u^\mu = \gamma(1,\boldsymbol{\beta})$ its normalized four-velocity, and $\tau = t/\gamma$ is the proper time.\\

The x-ray brilliance is closely related to the spectral angular density. Using the fact that, obviously, $\sigma_e \ll \sigma_L$ for a single electron,
\begin{equation}\label{eq:Bx_dndwdW}
   B_\mathrm{x} \simeq \frac{f N_e}{2\pi \sigma_e^2 }\omega_0\mathcal{S}(\omega_0,0).
\end{equation}

The dimensionless quantity $\omega_0\mathcal{S}(\omega_0,0)$ is the peak normalized spectral angular density, which is maximized for $\theta_\mathrm{x} = 0$ and therefore $\omega =\omega_0 = 2\gamma^2\omega_L (1-\beta\cos{\theta_L})$. In this section we will analyze how $\omega_0\mathcal{S}(\omega_0,0)$ depends on the interaction geometry of ICS. Specifically, how it depends on the laser pulse dimensions, interaction angle and electron energy.\\

The integral in Eq.~\eqref{eq:specangCovariant} is solved by considering a single electron that experiences the normalized vector potential, with amplitude $A(\tau)$, due the laser pulse. We assume that this normalized vector potential is Gaussian, \textit{i.e.} $A(\tau) = A_0\exp(-\tau^2/4\tau_{\mathrm{int}}^2 )$. Here, $\tau_{\mathrm{int}}$ is the proper interaction time of the electron with the laser pulse. Note that $\tau_{\mathrm{int}}$ is \textit{not} equal to the length of the laser pulse. The manner in which $\tau_{\mathrm{int}}$ depends on the parameters of the laser pulse is discussed later. $A_0$ can be written in terms of the laser pulse dimensions
\begin{equation}\label{eq:A0def}
        A_0^2 = \frac{3 \sigma_T   N_L}{2\alpha\pi^{3/2} }\frac{1}{\sigma_{L\parallel}\sigma_{L\perp} \omega_L\sigma_t}.
\end{equation}
Because of the asymmetry in the interaction geometry, we also have to consider an asymmetric laser pulse focus with two transverse sizes, $\sigma_{L\parallel}$ and $\sigma_{L\perp}$. Here, the subscripts $\parallel$ and $\perp$ refer to the directions parallel and perpendicular to the x-z plane, respectively.\\

When nonlinear effects can be neglected, \textit{i.e.} $A_0 \ll 1$, the peak normalized spectral angular density due to a single electron is given by the concise expression
\begin{align}
    \label{eq:dNdwdWmax_SingleElec}
  \omega_\mathrm{0}\mathcal{S}(\omega_\mathrm{0},0) &= \frac{1}{4 \pi}\alpha A_0^2\omega_\mathrm{0}^2\tau_\mathrm{int} ^2
\end{align}
For details, we refer to Appendix \ref{sec:appendixA}. In this expression all dependencies on the interaction geometry are implicit in $A_0$, $\omega_0$, and $\tau_\mathrm{int}$. The quantity  $\omega_0 \tau_\mathrm{int}$ plays an important role. It is equal to the number of oscillations that the electron makes during the interaction with the laser pulse, multiplied by the Lorentz factor $\gamma$. In either a head-on geometry ($\theta_L = \pi$) or a perfect co-propagating geometry ($\theta_L = 0$), $\omega_0\tau_\mathrm{int} = \gamma\omega_L \sigma_t$. For intermediate interaction angles ($0<\theta_L<\pi$), $\omega_0\tau_\mathrm{int}$ is reduced due to the less than infinite value of $\sigma_{L\parallel}$. \\

\begin{figure}
    \centering
    \includegraphics[width=1\linewidth]{ 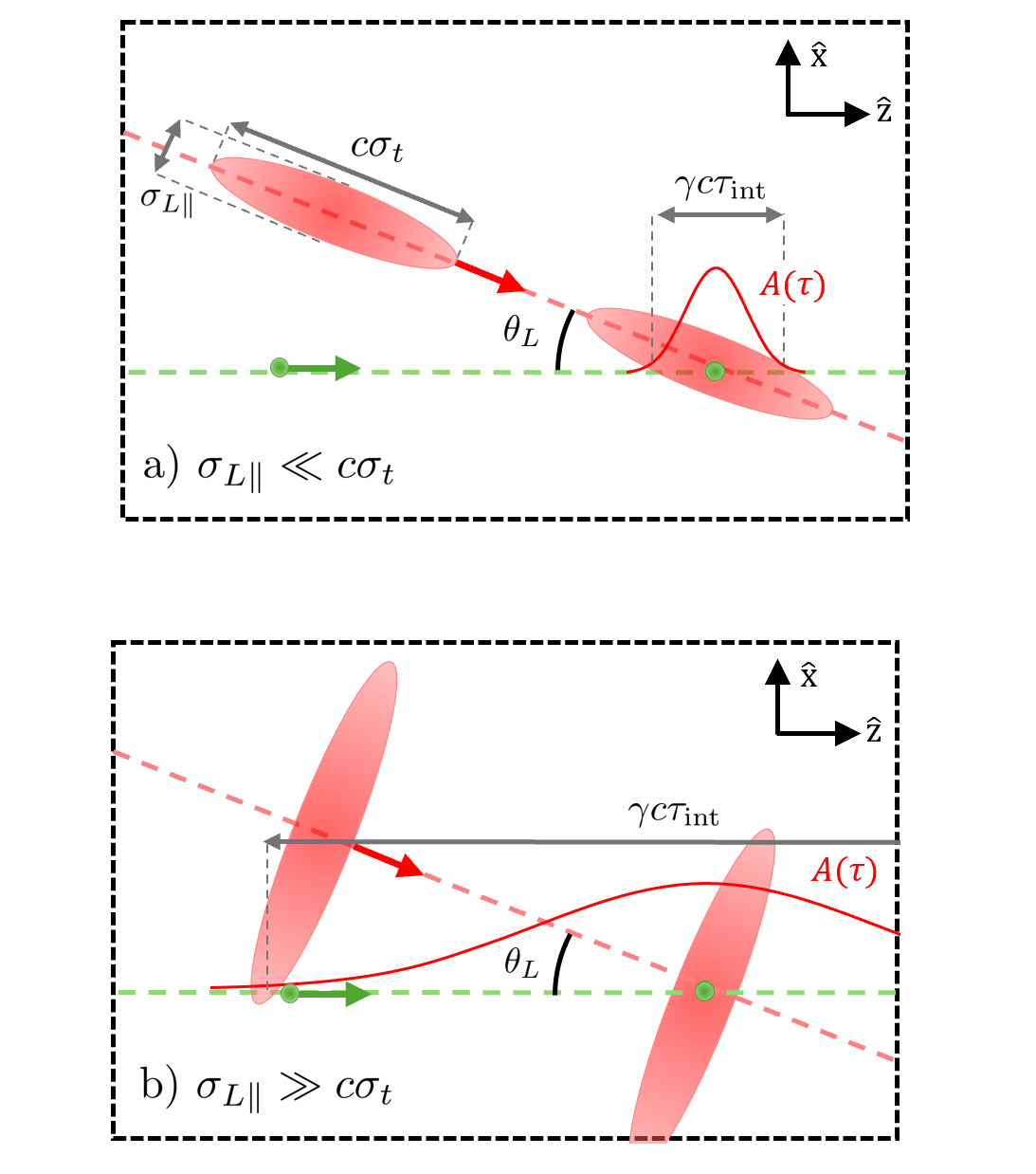}
    \caption{Illustrations on how the proper interaction time $\tau_\mathrm{int}$ is reduced by $\sigma_{L\parallel}$. a) show the case for $\sigma_{L\parallel}\ll c\sigma_t$ and the b) shows the case for $\sigma_{L\parallel}\gg c\sigma_t$ }
    
    \label{fig:ColinearGeometry}
\end{figure}
Figure ~\ref{fig:ColinearGeometry} illustrates how the overlap between the electron and the laser pulse is dependent on $\sigma_{L\parallel}$, for two cases: a) where $\sigma_{L\parallel}\ll c\sigma_t$; and b) where $\sigma_{L\parallel}\gg c\sigma_t$. The figure shows that the start of the interaction is not only determined by the laser pulse length, but  also by its transverse size $\sigma_{L\parallel}$. This is illustrated by the vector potential amplitude, as experienced by the electron (red solid curve). Figure~\ref{fig:ColinearGeometry} shows that $\tau_\mathrm{int}$ can become very large when $\theta_L\ll 1$ (as long as $\sigma_{L\parallel}\gg c\sigma_t$), which might infer increased x-ray yield. However, this is exactly canceled by the dependence of $\omega_0$ on $\theta_L$, when evaluating $\omega_0 \tau_\mathrm{int}$. This is because the total number of oscillations of the electron can never be larger than the number of oscillations in the laser pulse, $\omega_L\sigma_t$, which is Lorentz invariant.\\

%The transformation to the co-moving frame not only changes the interaction angle, the laser pulse also attains a pulse front tilt. For highly relativistic beams this pulse front tilt can be close to 90 degrees such that the roles of pulse length and transverse size are reversed. As a result, the interaction time becomes strongly dependent on $\sigma_{L\parallel}$.\\

The reduction in the interaction time can be quantified by a dimensionless parameter $\zeta \in [0,1]$, defined such that $\omega_0\tau_\mathrm{int} = \gamma \omega_L\sigma_t\zeta$, where 
\begin{align}\label{eq:zeta}
  \zeta \equiv \left[1 + \left(\frac{\beta \sin \theta_\mathrm{L}}{1-\beta\cos{\theta_L} }\frac{c \sigma_t}{\sigma_{L\parallel}}\right)^2\right]^{-1/2}.
\end{align}

\begin{figure}[h]
    \centering
    \includegraphics[width = \linewidth]{ 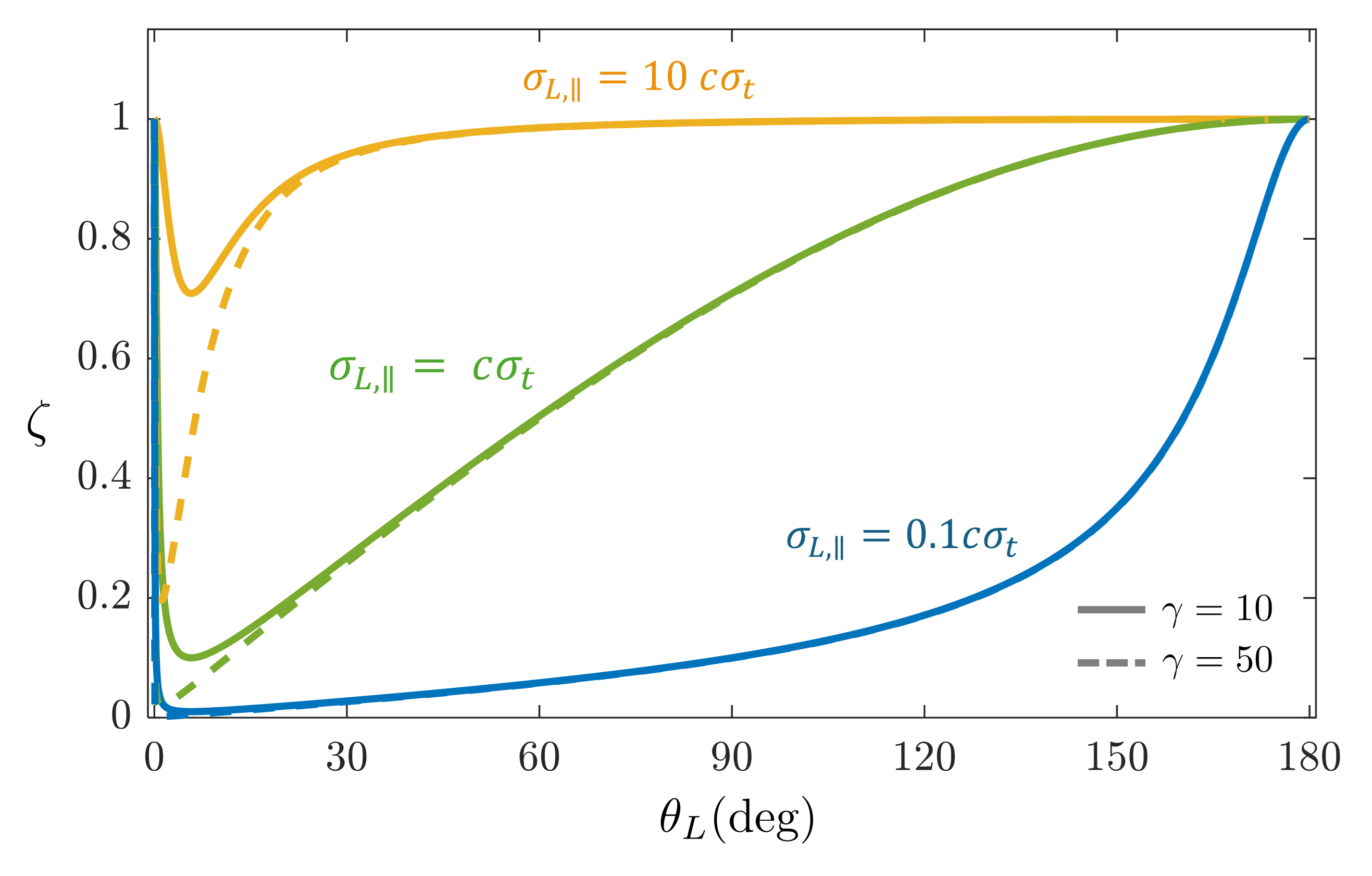}
    \caption{Reduction $\zeta$ of the overlap time as a function of the interaction angle $\theta_L$ for three values of the ratio $\sigma_{L\parallel}/c\sigma_t$ (blue, green, and yellow curves) for $\gamma = 10$ (solid curves) and for $\gamma = 50$ (dashed curves).}
    \label{fig:zetaplot}
\end{figure}

$\zeta$ describes the number of oscillations experienced by the electron, compared to the total number of oscillations in the laser pulse. The geometric reduction of the number of oscillations is determined by the ratio between the electron velocity component perpendicular to the laser pulse propagation ($\beta \sin\theta_L$) and the difference between the electron and laser velocity along the direction of the laser pulse propagation ($1-\beta\cos\theta_L$). The latter of which can become very small for small interaction angles. \\
Eq.~\eqref{eq:dNdwdWmax_SingleElec} can be rewritten such that the peak normalized spectral angular density becomes
\begin{equation}\label{eq:tau_int}
  \omega_\mathrm{0}\mathcal{S}(\omega_\mathrm{0},0) = \frac{1}{4 \pi}\alpha A_0^2\gamma^2\omega_L^2\sigma_t ^2\zeta^2.
\end{equation}
Figure~\ref{fig:zetaplot} shows how $\zeta$ depends on the interaction angle $\theta_L$ for various aspect ratios $\sigma_{L\parallel}/c\sigma_t$. In either the head-on geometry $(\theta_L = \pi)$, or a perfect co-propagating geometry ($\theta_L = 0$) the electron travels through all of the oscillations of the laser pulse and therefore $\zeta = 1$. However, for $0<\theta_L<\pi$, the electron experiences fewer oscillations, which strongly depends on $\theta_L$. This effect most extreme at small interaction angles, a \textit{grazing angle geometry}. $\zeta$ is minimum at $\theta_L = \cos^{-1}(\beta) \simeq  1/\gamma$, since at this angle the electron velocity is matched to the laser pulse velocity component along the z-axis.\\

The geometric reduction of the interaction time can be overcome by either applying a pulse front tilt to the laser that corrects for the tilted front in the co-moving frame \cite{steiniger2014optical,potylitsyn2023crab,schaap2025shallow}; or by utilizing a line focused laser pulse, where $\sigma_{L\parallel}\gg \sigma_{L\perp}$. In this work we will analyze the second solution.\\

As an example, a 100 fs laser pulse focused to $\sigma_L = 2.4 \ \mathrm{\mu m}$ interacting under $ \theta_L =10^\circ$ at $\gamma = 50$, lowers the interaction time by a factor $\zeta = 0.007$. As a result, the brilliance is reduced by more than four orders of magnitude. Utilizing a line focus, with aspect ratio $\sigma_{L\parallel}/\sigma_{L\perp} = 100$, we find that $\zeta = 0.58$. While the laser intensity is now lower (by a factor of 100), the x-ray brilliance is almost two orders of magnitude higher than if a tight symmetrical focus was applied.\\

The blue curve in Fig.~\ref{fig:zetaplot} shows that also for small deviations from $\theta_L = \pi$ the interaction time can become significantly reduced when utilizing a tight laser focus. This suggest that in a head-on geometry, it may pay off to approach $\theta_L = \pi$ as closely as possible, for instance by using parabolic mirrors with through holes \cite{zhang2024experimental}. \\

In Sec.~\ref{sec:optimization} we will find expressions for the optimized laser pulse dimensions. To find these, however, we first need to describe the effect of using a focused laser beam at arbitrary interaction angles. \\

%%%%%%%%%%%%%%%%%%%%%%%%%%%%%%%%%%%%%%%%%%%%%%%%
%%%%%%%%%%%%%%%%%%%%%%%%%%%%%%%%%%%%%%%%%%%%%%%%
\subsection{Single electron with a focused laser beam}\label{sec:colin_divergence}
Similar to Sec.~\ref{sec:divergence}, we have found that a tight laser focus is required to optimize the ICS interaction. Therefore we need to consider the effect of a focused laser pulse. In contrast to Sec.~\ref{sec:divergence}, we now only have a tight focus in the y-direction. Additionally, since the interaction angle is no longer parallel to the electron velocity, the electron no longer traverses the complete longitudinal profile of the laser pulse. The effect of a focused laser pulse for grazing angle ICS is analyzed using the substitution $A(\tau) \rightarrow A(\tau) /\sqrt{1+ \tau^2/2\tau_R^2}$, similar to the method presented in \cite{schaap2023ponderomotive}. Here we have introduced a proper Rayleigh time,
\begin{equation} \label{eq:tauR}   
    \tau_R = \frac{2\sqrt{2} k_L \sigma_{L,\perp}^2}{\gamma\beta c \cos{\theta_L}}\left(1+ \frac{\sigma_{L,\perp}^4}{\sigma_{L,\parallel}^4}\right)^{-1/2},
\end{equation}

which describes the proper time that the electron spends within the Rayleigh range of the laser pulse. When this time becomes small compared to the interaction time the number of scattered photons is reduced and the x-ray bandwidth is increased. The effect can be taken into account by the substitution

\begin{align}\label{eq:sigmatau_sub}
     \tau_\mathrm{int} \rightarrow  \tau_\mathrm{int}\psi\   \mathcal{K}_0 ({\psi^2/4})
\end{align}

with $\mathcal{K}_0 (x) = \exp(x) K_0(x)/\sqrt{2\pi}$ the exponentially scaled Bessel function of second kind and order zero, and $\psi = \tau_R/\tau_\mathrm{int}$. The role of $\psi$ is also similar to that of $u$ from Sec.~\ref{sec:divergence}. However, $\psi$ describes the effect of laser pulse divergence on the interaction time at any interaction angle. In a head-on geometry, $\psi = u $. 
This result is very similar of the result from Eq.~\eqref{eq:xrayTotDiv} since 
$\sqrt{\pi}\ \mathrm{erfcx}(x)\simeq\mathcal{K}_0(x^2/4)$.  \\

In a line focus geometry, where $\sigma_{L\perp}\ll\sigma_{L\parallel}$, $\psi$ can be rewritten as
\begin{equation}\label{eq:psi_linefocus}
    \psi \simeq \frac{1- \beta \cos{\theta_L}}{\beta\cos{\theta_L}}\frac{2 k_L^2 \sigma^2_{L\perp}}{\omega_L \sigma_t \zeta}
\end{equation}
When the interaction angle $\theta_L \ll 1$, $\psi \propto 1/\gamma^2$. This is a consequence of the long interaction length, $\gamma\beta c \tau_\mathrm{int}$, in a grazing angle geometry with a line focused laser pulse. Therefore, $\sigma_{L\perp}$ needs to be larger, compared to the head-on case, to ensure that the Rayleigh length in the perpendicular direction does not reduce the interaction time. In Sec.~\ref{sec:optimization} we will find an expression for $\sigma_{L\perp}$ that optimizes the x-ray brilliance.\\

In addition to a reduced interaction time, the angular spread of the focused laser pulse also induces an additional x-ray bandwidth that is not yet included in Eq.~\eqref{eq:sigmatau_sub}. The angular spread introduces a spread in the Doppler shift. This can be quantified by an expansion of Eq.~\eqref{eq:dopplerShift1} around $\theta_L$, resulting in  
\begin{equation}\label{eq:BW_Colin_lasdiv}
     \left.\sigma_{\omega}/\omega_\mathrm{x}\right|_{\sigma_{\theta L}} =   \frac{\beta \cos\theta_L}{1-\beta \cos{\theta_L}}      (\sigma_{\theta L \parallel}^2 +\sigma_{\theta L \perp}^2). 
\end{equation}
Here $\sigma_{\theta L\parallel}$ and $\sigma_{\theta L\perp}$ are angular spreads of the laser pulse parallel and perpendicular to the x-z plane respectively. We note that one would also expect a term in Eq.~\eqref{eq:BW_Colin_lasdiv} that is first order in $\sigma_{\theta L\parallel}$, however this therm is already taken into account by $\zeta$. For a line focused laser pulse, Eq.~\eqref{eq:BW_Colin_lasdiv} can we rewritten as
\begin{equation}\label{eq:BW_laserdiv_linefoc}
    \left.\sigma_{\omega}/\omega_\mathrm{x}\right|_{\sigma_{\theta L}} = 1/\left(  \omega_L \sigma_t \zeta \psi   \right).
\end{equation}

Eq.~\eqref{eq:dNdwdWmax_SingleElec} only takes into account the Fourier limited bandwidth
\begin{equation}\label{eq:BW_fourierlim_angle}
\left.\sigma_{\omega}/\omega_\mathrm{x}\right|_{\omega_L} = 1/\left[ 2 \omega_L \sigma_t \zeta \psi\  \mathcal{K}_0(\psi^2/4)  \right].
\end{equation}
By combining the above two expressions, we find the following expression for the peak normalized spectral angular density

\begin{align}\label{eq:dNdwdW_colin_diffraction}
      \omega_\mathrm{x} \mathcal{S}(\omega_\mathrm{x},0) &= \frac{1}{  \pi}\alpha A_0^2\gamma^2\omega_L^2\sigma_{t}^2\zeta^2\  \frac{ \psi^2 \ \mathcal{K}_0(\psi^2/4)^2 }{\sqrt{1+  4 \mathcal{K}_0(\psi^2/4)^2 }}
\end{align}
Other bandwidth contributions can be included in the same method as discussed in Sec.~\ref{sec:divergence}. For the laser parameters under consideration, the bandwidth due to nonlinear broadening can be neglected due to the use of the line focus.

%%%%%%%%%%%%%%%%%%%%%%%%%%%%%%%%%%%%%%%%%%%%%%%%
%%%%%%%%%%%%%%%%%%%%%%%%%%%%%%%%%%%%%%%%%%%%%%%%
\subsection{Focused electron and laser beams}
\label{sec:Focused electron and laser beams}
Up to now, we have only considered the interaction of a single electron with the laser pulse. We will now also take into account the finite electron bunch size and angular spread by means of correction terms. \\

The finite size of the electron beam can be taken into account by considering an electron with offset ${(x,y,z)}$ at $t= 0$. This electron will probe a lower intensity of the laser pulse and therefore radiate fewer x-rays. Note that the x-ray bandwidth remains unaffected. For an electron bunch with initial density distribution
\begin{equation}\label{eq:DensityIdeal}
    \rho_{e}(t = 0) = \frac{N_{e}}{(2\pi)^{3/2} \sigma_{e}^2\sigma_{z}}\exp\left(-\frac{x^2 + y^2}{2\sigma_{e}^2} -\frac{z^2}{2\sigma_{z}^2}\right),
\end{equation}
we find that the total number of scattered x-rays is reduced by a factor $1 + \Sigma$, where 
\begin{align}\label{eq:Sigma2}
\begin{split}
     \Sigma =\   & \frac{\sigma_e^2( 1- \beta\cos{\theta_L} - \sin^2{\theta_L})^2/\beta^2  + \sigma_z^2\sin^2{\theta_L}}{4 \sigma_{L\parallel}^2 (1-\beta\cos{\theta_L})^2 +   4 \beta^2 c^2 \sigma_t^2  \sin^2{\theta_L}}  \\
 &+ \frac{\sigma_{e}^2}{4\sigma_{L\perp}^2}.
\end{split}
\end{align}
In a head-on geometry, this expression reduces to
\begin{equation}
     \Sigma \underset{\theta_L= \pi}{=}\frac{\sigma_{e}^2}{4 \sigma_{L\parallel}^2} +\frac{\sigma_{e}^2}{4 \sigma_{L\perp}^2} .
\end{equation}
By expanding Eq.~\eqref{eq:Sigma2} around $\theta_L = 0$ and assuming that $\sigma_{L\perp} \ll \sigma_{L\parallel}$, we find that 
\begin{equation}
    \Sigma \underset{\theta_L \ll 1 }{\simeq}\frac{\sigma_{e}^2}{4 \sigma_{L\perp}^2} + \frac{\sigma_z^2 }{\sigma_{L\parallel}^2}\gamma^4\theta_L^2
\end{equation}

The important result here is that the electron pulse length, $\sigma_z$, starts to have a significant influence on the total x-ray flux compared to the head-on case. Note that $\gamma \theta_L \gg 1$, even in a grazing angle geometry (See Eq.~\eqref{Eq:DopplerGrazing}). \\

Besides the well known bandwidth contribution of the electron angular spread due to the spread in emission angle, $\left.\sigma_{\omega}/\omega_\mathrm{x}\right|_{\sigma_{\theta \mathrm{x}}} = \gamma^2\sigma_{\theta e}^2 $, in a grazing angle geometry we also have to take into account that the electron angular spread changes the interaction angle. This creates an x-ray bandwidth contribution
\begin{equation}\label{eq:BW_Colin_electrondiv}
      \left. \Delta \omega/{\omega}\right|_{\sigma_{\theta e}} =  \frac{\beta \sin\theta_L}{1-\beta \cos{\theta_L}}  \sigma_{\theta e} \simeq \sqrt{\frac{\omega_\mathrm{x}}{\omega_L}}2\gamma \sigma_{\theta e}.
\end{equation}
The approximation on the right holds for $\theta_L \ll 1$.\\

Finally, we must take into account the effect of the interaction length, $\gamma \beta c\tau_\mathrm{int}$, becoming longer than the electron beta function, $\hat{\beta} = \epsilon_n/(\gamma \sigma_{\theta e}^2)$. In this regime the x-ray source size varies during the interaction and we can no longer use the electron beam size in the focus as the x-ray source size. Instead we have to use the effective x-ray source size, obtained by a laser intensity-weighted average of the electron beam size:
\begin{equation}\label{eq:effective_sige}
    \sigma_{e\mathrm{,eff}}^2 = \sigma_e^2 \left[1 + \frac{1}{2}\left(\gamma^2\sigma_{\theta e}^2 c \tau_\mathrm{int}/{\epsilon_n}\right)^2\right].
\end{equation}
Equation \eqref{eq:effective_sige} should be substituted in Eq.~\eqref{eq:Bx_dndwdW}. It is not required to apply the substitution in Eq.~\eqref{eq:Sigma2} since we will find that in the above mentioned regime, the electron beam becomes very small compared to the laser focus. 

%%%%%%%%%%%%%%%%%%%%%%%%%%%%%%%%%%%%%%%%%%%%%%%%
%%%%%%%%%%%%%%%%%%%%%%%%%%%%%%%%%%%%%%%%%%%%%%%%
\subsection{Optimization of a grazing angle interaction geometry}\label{sec:optimization}
All ingredients which are required for optimizing the grazing angle interaction geometry are now in place. To arrive at closed form analytic expressions, we first consider the interaction of a single electron with a focused laser pulse and optimize the peak normalized spectral angular density. Next we add the contributions for a realistic electron beam to calculate the corresponding x-ray brilliance. As a final step, we investigate the behavior for small interaction angles $\theta_L \ll 1$, to find the optimized electron energy. This method yields analytical expressions for the laser pulse dimensions, interaction angles, and electron energy for a desired x-ray energy.\\

For the interaction of a single electron, we optimize the peak normalized spectral angular density given by Eq.~\eqref{eq:dNdwdW_colin_diffraction}, to find optimized values of $\sigma_{L\parallel}$ and $\sigma_{L\perp}$. The dependency on the laser pulse dimensions is implicit in $A_0$ and $\zeta$, given by \cref{eq:A0def,eq:zeta} respectively. We now consider a line focused geometry such that $\psi$ is given by Eq.~\eqref{eq:psi_linefocus}, and recognize that $A_0^2 \propto 1/\sigma_{L\parallel}\sigma_{L\perp}$. We now find that the peak normalized spectral angular density scales as 
\begin{equation}\label{eq:h(psi)}
\begin{split}
   \omega_\mathrm{x} \mathcal{S}(\omega_\mathrm{x},0) \propto  \psi^{3/2}  
h(\psi)\ \zeta^{3/2}/\sigma_{L\parallel} \\
   \mathrm{with}\  h(\psi) = \frac{  \mathcal{K}_0(\psi^2/4)^2 }{\sqrt{1+  \mathcal{K}_0(\psi^2/4)^2 }}.
\end{split}
\end{equation}

The optimized $\sigma_{L\parallel}$ is found by maximizing $ \zeta^{3/2}/\sigma_{L\parallel}$, which occurs when
\begin{equation}\label{eq:sigL_opt2}
    \sigma_{L\parallel}^{\mathrm{opt}} = \frac{1}{\sqrt2}\frac{\sin\theta_L}{1-\beta \cos{\theta_L}}\beta c \sigma_t.
\end{equation}
Under this condition, $\zeta = \sqrt{1/3}$. Next, to find the optimized $\sigma_{L\perp}$, we optimize $\psi^{3/2}h(\psi)$, which has a maximum at $\psi_{\mathrm{max}} = 1.91$. When taking into account additional bandwidth contributions such as the electron energy spread, resulting additional terms in the denominator of $h(\psi)$. This will have the effect of decreasing $\psi_{\mathrm{max}}$. \\

Substituting $\zeta = \sqrt{1/3}$ and $\psi = \psi_{\mathrm{max}}$ into Eq.~\eqref{eq:psi_linefocus}, we also find an expression for the optimized $\sigma_{L\perp}$:
\begin{equation}\label{eq:sigL_perp_opt2}
    \sigma_{L\perp}^{\mathrm{opt}} = \frac{1}{2} \sqrt{\frac{ c \sigma_t}{ k_L} \left|\frac{\beta\cos{\theta_L}}{1-\beta\cos{\theta_L}}\right|\sqrt\frac{1}{3}\psi_{\mathrm{max}}}.
\end{equation}
The absolute value signs are required since the argument becomes negative when $\theta_L > 90^\circ$.\\

\begin{figure}
    \centering
     \includegraphics[width=\linewidth]{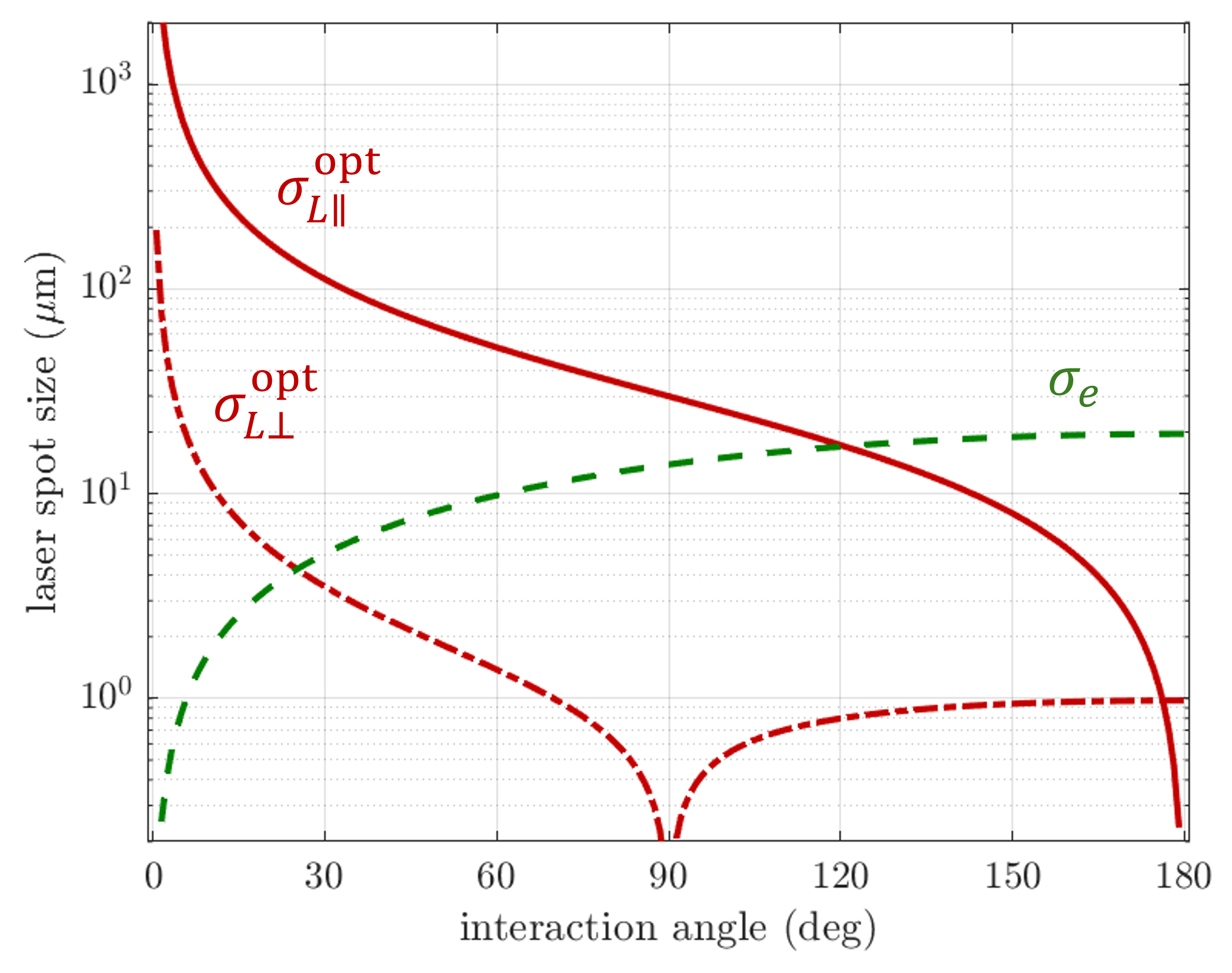}
    \caption{Optimized laser spot sizes $\sigma_{L\parallel}^{\mathrm{opt}}$and $\sigma_{L\perp}^{\mathrm{opt}}$ (red solid and dash-dotted curves) and electron spot size $\sigma_e$ (green dashed curve) for the generation of 500 eV x-rays as a function of interaction angle. }
    \label{fig:laserspotsizes}
\end{figure}

The red curves in Fig.~\ref{fig:laserspotsizes} show the optimized laser spot sizes $\sigma_{L\parallel}$ and  $\sigma_{L\perp}$ for the generation of 500 eV x-rays under different interaction angles, according to \cref{eq:sigL_opt2,eq:sigL_perp_opt2}. The figure shows that for small interaction angles the optimal laser spot size $\sigma_L$ becomes large in both directions; however, the line focus is still maintained: $\sigma_{L\parallel}\gg\sigma_{L\perp}$ over the entire angular range. The green dashed curve shows $\sigma_e$ for $\epsilon_n =$ 200 nmrad and $\sigma_\theta$ = 1 mrad. Since the optimized laser spot size can become very small outside of the grazing angle geometry, $\sigma_e$ should always be used as a lower bound for $\sigma_L$.  \\

The x-ray brilliance can be calculated by including the effects of the finite electron beam and effective spot size into Eq.~\eqref{eq:Bx_dndwdW}, resulting in
\begin{equation}\label{eq:Bx_dndwdW_3D}
    B_\mathrm{x} = \frac{fN_e}{2 \pi \sigma_e^2 (1+\kappa^2/2)}\frac{1}{1+\Sigma} \omega_\mathrm{x} \mathcal{S}(\omega_\mathrm{x},0).
\end{equation}
Here, we have introduced $\kappa~=~\gamma^2 \sigma_{\theta e}^2 c \tau_\mathrm{int} /\epsilon_n$ which corrects for the effective spot size over the interaction length. For arbitrary interaction angles and beam dimensions the parameters $\zeta,\psi$ and $\Sigma$ should be calculated using \cref{eq:zeta,eq:tauR,eq:Sigma2},  respectively. However, in the optimized grazing angle geometry, $\zeta = \sqrt{1/3}$ , $\psi = 1.91$, and $\Sigma$ becomes
\begin{equation}
    \Sigma = \frac{\sigma_z^2}{6\beta^2 c^2\sigma_t^2}.
\end{equation}
\begin{figure}
    \centering
     \includegraphics[width=\linewidth]{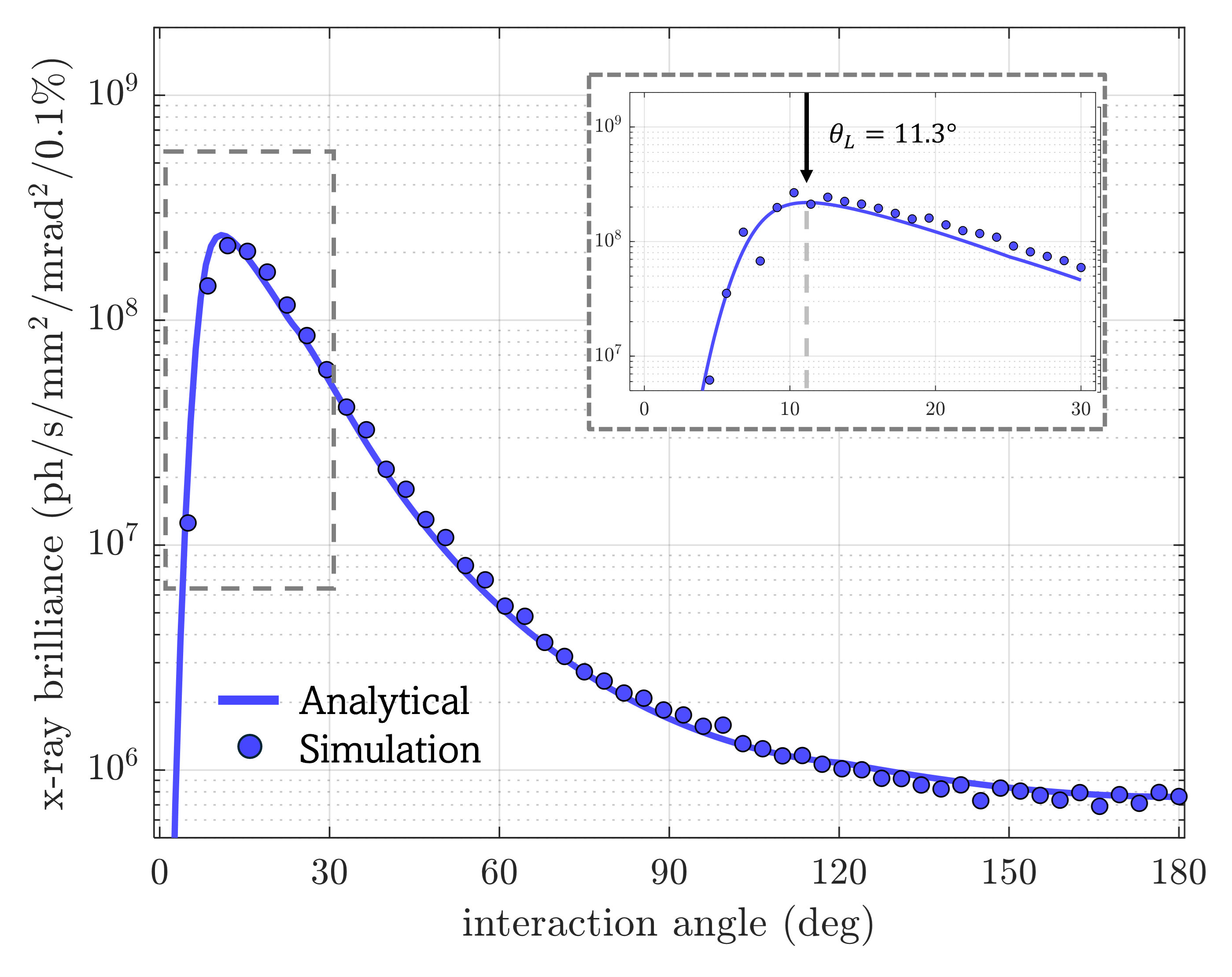}
    \caption{X-ray brilliance as a function of interaction angle for the generation of 500 eV x-rays, using optimized laser spot sizes $\sigma_{L\parallel}^{\mathrm{opt}}$ and $\sigma_{L\perp}^{\mathrm{opt}}$. The inset zooms in on the $0^{\circ}-30^{\circ}$ angular range, where the arrow denotes the interaction angle corresponding to the optimized electron energy $\gamma_\mathrm{opt}$ }
    \label{fig:brillianceAngle}
\end{figure}

In Fig.~\ref{fig:brillianceAngle} x-ray brilliance for the generation of 500 eV x-rays is plotted as a function of interaction angle. While the interaction angle is varied, the laser pulse dimensions are also changed, according to Eqs.~\eqref{eq:sigL_opt2} and \eqref{eq:sigL_perp_opt2}. The normalized emittance of the electron beam is $\epsilon_n =$ 200 nmrad and the electron beam angular spread $\sigma_\theta=1$~mrad. The electron energy is determined by the interaction angle and the fixed x-ray energy of 500 eV. Here $\sigma_e$ is used as a lower bound for $\sigma_{L \parallel}$ and $\sigma_{L \perp}$.\\

The figure clearly shows how brilliance increases with decreasing interaction angles, but only up to a point. At small interaction angles, the effective source size becomes so large that the brilliance drops abruptly. This results in an optimal interaction angle, or equivalent electron energy. To find an expression for this optimized electron energy, we analyze how $\sigma_{L\parallel}^{\mathrm{opt}}$ and $\sigma_{L\perp}^{\mathrm{opt}}$ scale for large electron energy. By series expansion of Eq.~\eqref{eq:dopplerShift1} for $\theta_L \ll 1$ we find that
\begin{align}
     \sigma_{L\parallel}^{\mathrm{opt}} &\simeq  \sqrt{\frac{\omega_L}{\omega_\mathrm{x}}}\ 2\gamma  c \sigma_t,\\
     \sigma_{L\perp}^{\mathrm{opt}} &\simeq  \sqrt{\frac{\psi_{\mathrm{max}}}{  \omega_\mathrm{x}\sigma_t} \sqrt\frac{1}{3} }\  \gamma c \sigma_t.   
\end{align}
The aspect ratio of the optimized line focus is therefore given by  $\sigma_{L\parallel}/\sigma_{L\perp}\simeq 2.15\sqrt{\omega_L\sigma_t/\psi_{\mathrm{max}}}$. Since the laser spot sizes in both directions become linear with the electron energy, the peak normalized spectral angular density becomes independent of $\gamma$. As a consequence, the x-ray brilliance scales according to 
\begin{equation}
    B_{\mathrm{x}} \propto\frac{1}{\sigma_{e\mathrm{,eff}}^2 }=\frac{ \sigma_{\theta e}^2}{\epsilon_n^2 }\frac{\gamma^2}{1+ \kappa^2/2}.
\end{equation}
Since $\kappa \propto\gamma^3$, we can find the electron energy, $\gamma_{\mathrm{opt}}$, that optimizes the x-ray brilliance:
\begin{equation}\label{eq:gammaopt}
   \gamma_{\mathrm{opt}}  = \left[ \frac{\omega_L\sigma_{\theta e}^2}{\omega_\mathrm{x} \epsilon_n} 4 c \sigma_t \zeta \psi\ \mathcal{K}_0(\psi^2/4)\right]^{-1/3}. 
\end{equation}

Combining Eq.~\eqref{eq:gammaopt} with Eq.~\eqref{eq:dopplerShift1} thecorresponding interaction angle is obtained. As an example for the parameters used in Fig.~\ref{fig:brillianceAngle}, the optimized electron energy $\gamma = 103$ and the corresponding optimized interaction angle $\theta_L = 11.3^\circ$. This is in good agreement with the result shown in Fig.~\ref{fig:brillianceAngle}.\\

Finally, Eqs.~\eqref{eq:sigL_opt2} and \eqref{eq:sigL_perp_opt2} can be used to find the optimized laser pulse dimensions, completing the full characterization of the optimized grazing angle interaction geometry.

\subsection{Performance of a grazing angle soft x-ray Compton source}\label{sec:performance}
As an example, we apply the theoretical framework described above to the design of a soft x-ray Compton source with continuous tunability between 0.25~keV and 2.0~keV. This energy range is highly relevant for life sciences \cite{kordel2020laboratory,weinhardt2025soft} and catalysis \cite{cao2020emerging,beaumont2020soft}. In this energy range, achromatic reflective x-ray optics with focal lengths less than 1~m are available \cite{kimura2022soft}, which could keep the footprint within the scale of a university lab. \\

\begin{figure}
    \centering
    \includegraphics[width = \linewidth]{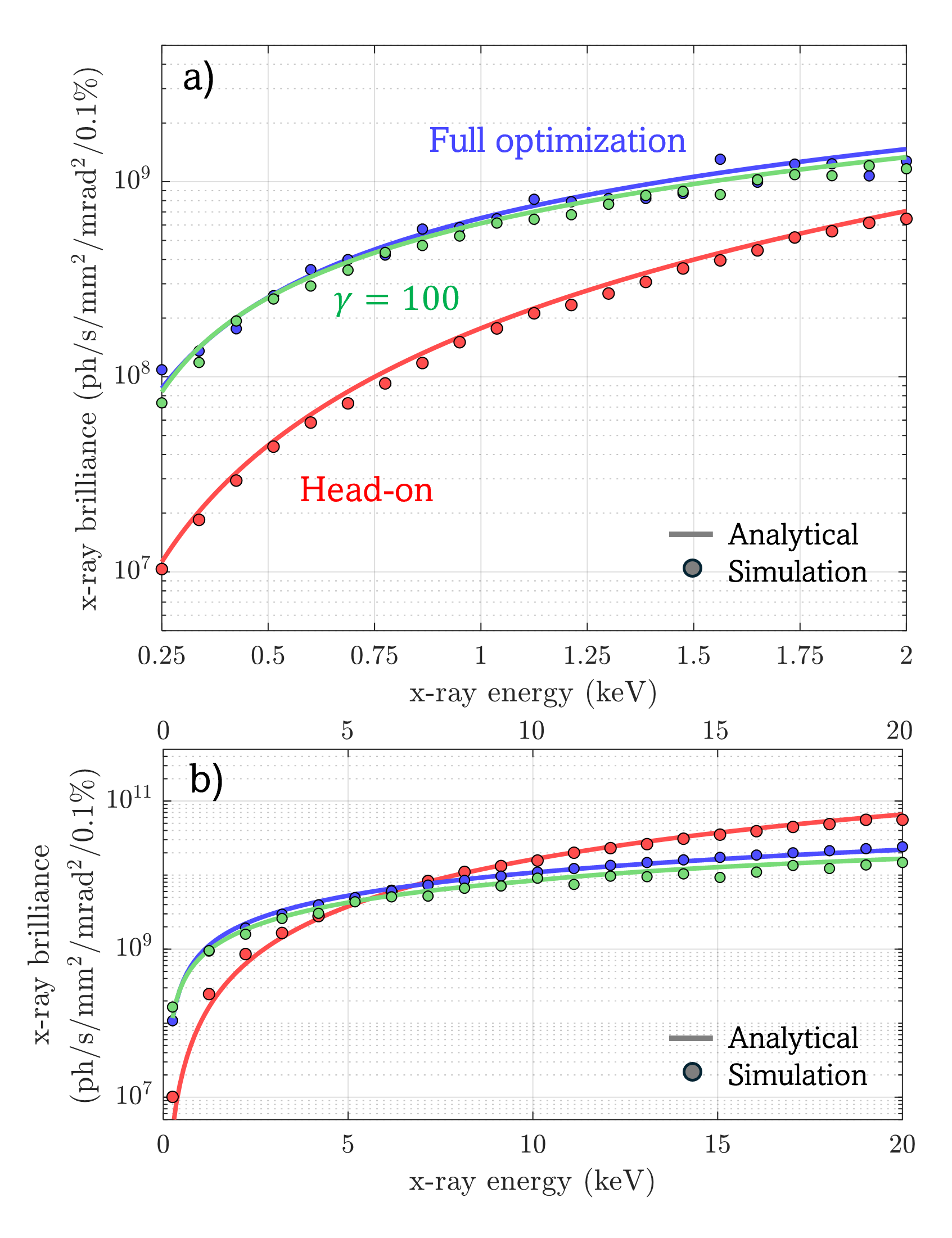}
    \caption{X-ray brilliance as a function of x-ray energy for three different optimized geometries: head-on scattering (red); grazing angle scattering with optimization of both the interaction angle and the electron energy (blue); and grazing angle scattering with the electron energy kept fixed at $\gamma = 100$ (green). (a) Results for soft x-rays in the range 0.25-2~keV and (b) for the entire range of 0-20~keV. }
    \label{fig:sourceperformance}
\end{figure}

The electron beamline is based on a high repetition rate C-band photoinjector \cite{lucas2023toward,alesini2026design}. This injector creates $Q = 200$ pC electron bunches with a normalized emittance of $\epsilon_n = 200$ nmrad $ (\eta = 14\mathrm{\ nmrad/\sqrt{pC}}) $ and 0.2\% energy spread at a repetition rate of $f = 1 $~kHz. Subsequent LINACs can be used to accelerate the electrons to the required energy.  The electrons are focussed to the interaction point with an angular spread of $\sigma_{\theta e} = 1$ mrad. The interaction laser is a 100 fs, 5mJ, Yb-fiber system with a central wavelength $\lambda_L = 2\pi c/\omega_L = 1030 {\,\rm nm}$, also operating at 1 kHz repetition rate \cite{LightConversion_Pharos}. We assume the electron pulse length to be equal to the laser pulse length. \\

We compare the x-ray source performance for three geometries. First, a full optimization as described in the previous section is performed, where the electron energy, the interaction angle, and the laser pulse shape are optimized. Second, an optimization is performed in which the electron energy is kept constant and only the interaction angle and the laser pulse shape are optimized. This has the advantage that no re-alignment of the electron optics is required, while the x-ray performance is hardly affected. Finally, a head-on geometry is optimized as described in Sec.~\ref{sec:head-on ics}. The resulting analytically optimized x-ray brilliances in the three different geometries are plotted with solid curves in Fig.~\ref{fig:sourceperformance}a for the $0.25-2$~keV soft x-ray energy range and in Fig.~\ref{fig:sourceperformance}b for the entire $0-20$~keV x-ray energy range. Also shown are the results of charged particle tracking simulations (General Particle Tracer \cite{PulsarPhysicsGeneralGPT}) combined with the numerical calculation of the Liénard-Wiechert potentials \cite{thomas2010algorithm}, which are indicated by dots. The analytical model results are in excellent agreement with the numerical calculations.\\

We find that the grazing angle geometry is able to improve the brilliance of an ICS source in the soft x-ray regime by almost an order of magnitude, compared to the head-on geometry. In addition, we find that the optimization of the grazing angle geometry is hardly affected if the electron energy is not optimized and kept fixed at $\gamma = 100$. Figure~\ref{fig:sourceperformance}b shows that the optimized head-on geometry outperforms the grazing angle geometry only at x-ray energies exceeding 7 keV for the parameters considered.\\

The performance of the grazing angle geometry could be improved even further by using an appropriate pulse front tilt such that $\zeta \rightarrow 1$. In this case, the x-ray brilliance would increase by approximately the aspect ratio of the laser pulse focus $\sigma_{L\parallel}/\sigma_{L\perp}\simeq 2.15\sqrt{\omega_L\sigma_t/\psi_{\mathrm{max}}}$.\\

Additionally, the performance of the head-on geometry can be increased by using an appropriately chirped laser pulse to eliminate the effect of nonlinear broadening, which would reduce $\chi \rightarrow 0$.
%%%%%%%%%%%%%%%%%%%%%%%%%%%%%%%%%%%%%%%%%%%%%%%%
%%%%%%%%%%%%%%%%%%%%%%%%%%%%%%%%%%%%%%%%%%%%%%%%
\section{Conclusion}
In this paper, we have presented a theoretical approach to optimize inverse Compton scattering x-ray sources. We have found closed form analytical expressions to describe the x-ray beam properties in terms of the laser and electron beam parameters. We find that the interaction angle is an important optimization parameter, which has not yet been treated in great detail. The model not only provides physical intuition on the important scaling of parameters involved in Compton scattering x-ray generation, but also allows for analytical optimization, which is usually done by time consuming simulations. The presented method can serve as a tool for the design of future Compton scattering x-ray sources and the optimization of existing inverse Compton scattering beamlines.\\

We find that, apart from the conventional head-on interaction geometry, a co-propagating grazing angle interaction can be particularly effective for generating soft x-rays. In general, the optimization involves a subtle interplay between several beam parameters. For a head-on geometry we find the surprising result that lowering the bunch charge can lead to a higher x-ray brilliance while maintaining the same x-ray flux. For soft x-ray wavelengths, optimization of a co-propagating grazing angle geometry results in an order of magnitude higher x-ray brilliance, compared to the optimized head-on geometry.\\

Even higher x-ray brilliance could be achieved in a head-on geometry by using chirped laser pulses to suppress the nonlinear broadening. The grazing angle geometry could be optimized even further by employing a laser pulse front tilt, which increases the interaction time without loss of intensity.\\

\begin{acknowledgments}
We thank Brian Schaap for various fruitful discussions.\\

This research was funded by EU Interreg Vlaanderen-Nederland (Smart*Light project 2.0), by the Ministry of Economic Affairs in the Netherlands through a TKI-grant, and by the Dutch Research Council (NWO) in The Netherlands, through the Industrial Partnership Program (IPP) grant 'ColdLight: From laser-cooled atoms to coherent soft X-rays'.
\end{acknowledgments}

\appendix 
\section{Derivation of the proper interaction time}\label{sec:appendixA}
We describe ICS in the framework of covariant electrodynamics, when  electron recoil  be ignored ( $4\gamma^2(1-\cos\theta_L)\hbar\omega_L/\gamma mc^2\ll 1$ ), by assuming an electron with normalized 4-velocity $u^{\mu}$ and 4-trajectory $x^{\mu}$, interacting with a laser pulse described by the normalized vector potential 

\begin{equation}
    A^{\mu}(\tau) = A_0 \exp{(-\tau^2/4\tau_\mathrm{int}^2)}\exp{[i\phi(\tau)]}\epsilon^\mu,
\end{equation}

with $\phi(\tau) = k_L^\nu x_\nu(\tau)$ the laser phase and $\epsilon^\mu$ is the laser 4-polarization. It is important to note that $A^\mu$ describes the vector potential as experienced by the electron, i.e. evaluated at the position and time described by the electron trajectory $x^\mu$. To arrive at Eq.~\eqref{eq:dNdwdWmax_SingleElec} we assume that the amplitude vector potential remains small $(A_0\ll1)$ and that the laser pulse envelope varies slowly compared to the laser phase. Additionally, we assume that the laser polarization is perpendicular to the initial electron 4-velocity $u_0^\mu$. The spectral angular density of the radiated field is given by \cite{schaap2022photon}

\begin{widetext}
    \begin{equation}\label{eq:App:SpecAngDensity}
\begin{split}
    \mathcal{S}(\omega,\theta_\mathrm{x}) = \frac{
    \alpha A_0^2}{4\pi}\omega\tau_\mathrm{int}^2 \exp\left[-\gamma^2\tau_\mathrm{int}^2(1-\beta\cos\theta_\mathrm{x})^2\left(\omega-\omega_L\frac{1-\beta\cos\theta_L}{1-\beta\cos\theta_\mathrm{x}}\right)^2\right] \left[1- \left(\frac{\sin\theta_\mathrm{x}\sin\varphi_\mathrm{x}}{\gamma(1-\beta\cos\theta_\mathrm{x})
   }\right)^2\right],
\end{split}
\end{equation}
\end{widetext}

with $\varphi_\mathrm{x}$ the polar angle of the emitted radiation. In the exponent we can directly identify the angular dependent Doppler shift of Eq.~\eqref{eq:dopplerShift1} and the Fourier limited bandwidth $\sigma_{\omega}/\omega_\mathrm{x} = \left[2 \gamma(1-\beta\cos{\theta_L})\omega_L\tau_\mathrm{int}\right]^{-1}$. Evaluation of Eq.~\eqref{eq:App:SpecAngDensity} at $\omega = \omega_\mathrm{x}$ and $\theta_x = 0$ results in Eq.~\eqref{eq:dNdwdWmax_SingleElec}. By integration over the x-ray frequency we find the angular radiation distribution 

\begin{equation}
     \frac{\partial N_\mathrm{x}}{\partial\Omega} = \frac{\alpha A_0^2 }{2\sqrt{2\pi}}\frac{\omega_\mathrm{x} \tau_\mathrm{int} }{1-\beta\cos{\theta_\mathrm{x}}}\left[1- \left(\frac{\sin\theta_\mathrm{x}\sin\varphi_\mathrm{x}}{\gamma(1-\beta\cos\theta_\mathrm{x})}\right)^2\right],
\end{equation}

from which we can derive the characteristic $1/\gamma$ cone. When $u_0^\nu\epsilon_\nu \neq 0$, this distribution changes. However, if $\gamma \theta_L \gg1$ this amounts to a rotation along $\varphi_\mathrm{x}$.\\

\begin{figure}
    \centering
    \includegraphics[width=0.8\linewidth]{ 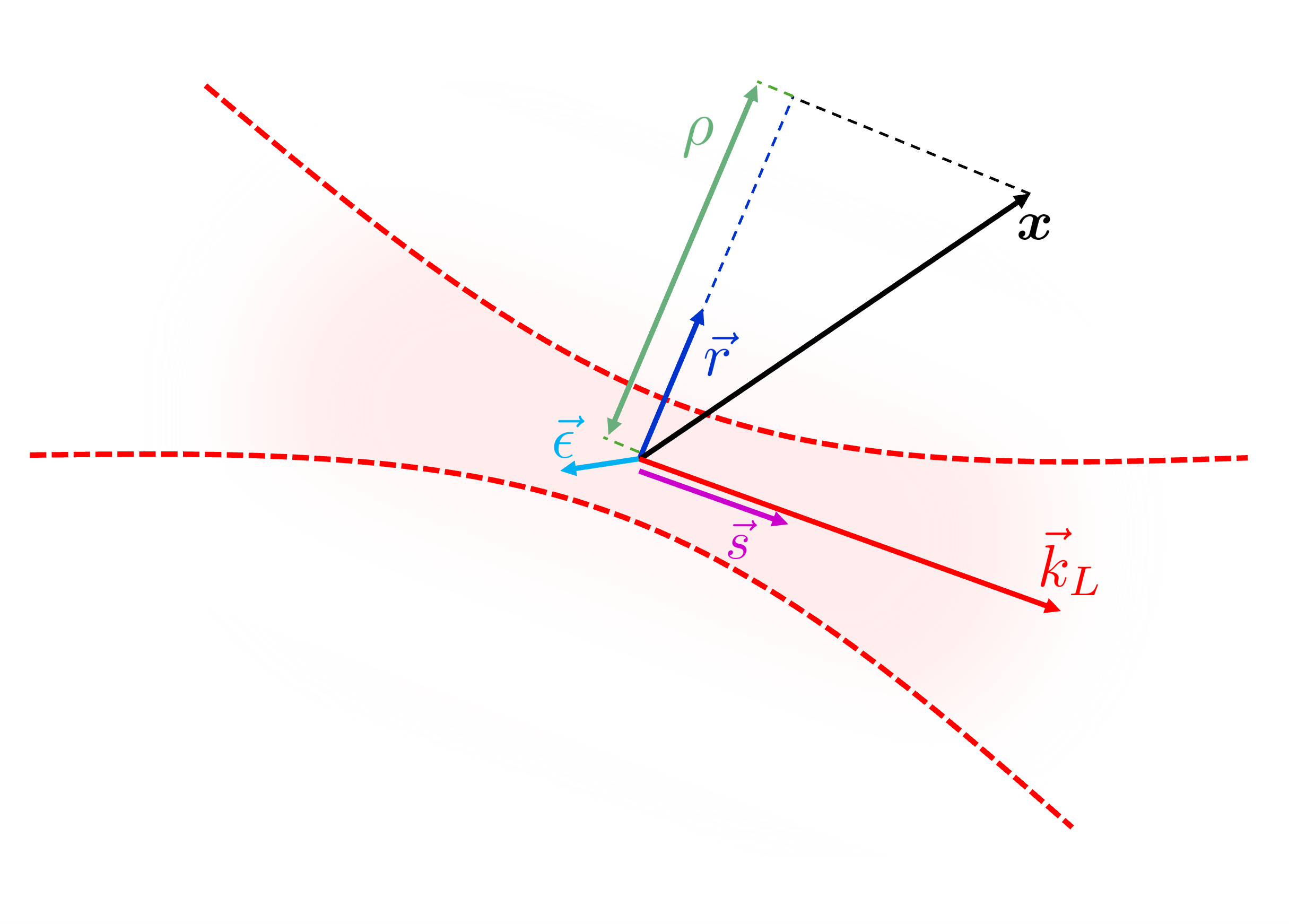}
    \caption{Illustration of the spatial pats of $k_L^\mu, r^\mu, \epsilon^\mu$ and $s^\mu$. The red curve represents the envelope of a Gaussian laser pulse.}
    \label{fig:3DLaserPulse_Vectors}
\end{figure}

To find the proper interaction time $\tau_\mathrm{int}$ we must describe the 3D laser pulse in Lorentz invariant quantities. For this we must introduce the following 4-vectors and corresponding invariant contractions:
\begin{align*}
 \phi =&\  x_{\nu}k^{\nu}_L &  k_\mathrm{L}^{\mu} =&\  k(1,\boldsymbol{\hat{n}}) \\
  \rho =&\  x_{\nu}r^{\nu} & r^{\mu} =&\  (0,\boldsymbol{\hat{\epsilon}} \times \boldsymbol{\hat{n}})  \\
  \eta =&\  x_{\nu}\epsilon^{\nu}   & \epsilon^{\mu} =&\  (0,\boldsymbol{\hat{\epsilon}})\\
  \varsigma =&\  x_{\nu}s^{\nu}   & s^{\mu} =&\  (0,\boldsymbol{\hat{n}})
\end{align*}
Figure \ref{fig:3DLaserPulse_Vectors} illustrates how the spatial components of these 4-vectors relate to the laser pulse.
We note that, while $r^\mu$ and $s^\mu$ can be constructed in any reference frame, different observers may not construct the same 4-vectors. These vectors should therefore be interpreted as a tool to describe a laser pulse, rather than manifestly covariant 4-vectors.\\

We can now describe the vector potential as
\begin{equation}
\begin{split}
           A^\mu = A_0\exp{\left[-\frac{\rho(\tau)^2}{4 \sigma_{L\parallel}^2}- \frac{\eta(\tau)^2}{4 \sigma_{L\perp}^2}-\frac{\phi(\tau)^2}{4 \omega_L^2\sigma_t^2}   \right]}\times\\
           \exp{[i\phi(\tau)]}\epsilon^\mu.
\end{split}
\end{equation}
We now assume that the electron has a ballistic trajectory, $x^\mu \simeq u_0^\mu c \tau$, where defined that $x_0^\mu = 0$. This allows the vector potential to be rewritten as
\begin{equation}\label{eq:GaussianPulseCovariant}
\begin{split}
           A^\mu = A_0\exp{\left(-\left[\frac{(u_0^\nu r_\nu)^2}{4 \sigma_{L\parallel}^2}+\frac{(u_0^\nu k_{L\nu})^2}{4 \omega_L^2\sigma_t^2}\right] c^2 \tau^2   \right)}\times\\
           \exp{[i\phi(\tau)]}\epsilon^\mu.
\end{split}
\end{equation}
Using $u_0^\nu r_\nu = \gamma \beta \sin{\theta_L}$ and $u_0^\nu k_{L\nu} = \gamma k_L (1-\beta \cos{\theta_L})$ the expression for the proper interaction time (Eq.~\eqref{eq:tau_int}) can be identified. To incorporate the finite electron beam waist, we take $x_0^\mu = ({0,x_0,y_0,z_0})$ and integrate over the electron distribution.\\

Equation \eqref{eq:GaussianPulseCovariant} does not describe a true Gaussian $\mathrm{TEM}_{00}$ mode. Importantly it lacks a description of the divergence, which can be corrected for using the substitution
\begin{equation}
    A^\mu \rightarrow A^\mu/\sqrt{1 + (\varsigma/z_R)^2}.
\end{equation}
Using $\varsigma = u_0^\nu s_\nu c \tau = \gamma\beta\cos{\theta_L} c \tau$ we can identify the standard integral \cite{abramowitz1948handbook}

\begin{equation}
    \int d\tau  \frac{\exp\left(-\tau^2/4 \tau_\mathrm{int}^2\right)}{\sqrt{1+\tau^2/2\tau_R^2}} = \tau_\mathrm{int} \psi\exp{(\psi^2/4)}K_0(\psi^2/4),
\end{equation}

with $\psi = \tau_R/\tau_\mathrm{int}$. Comparing this to the Gaussian standard integral results in Eq.~\eqref{eq:sigmatau_sub}.\\
\section{Optimization of other x-ray quantities}\label{app:quantities}
The optimization presented is this paper focuses on the average x-ray brilliance only. However, other x-ray quantities such as the spectral density, brightness or flux, could be the relevant quantity to optimize depending on the application. Here we show how the presented framework can be optimized for different quantities than the x-ray brilliance. Here it is discussed for the head-on case (Section II), however the same method can be applied to the framework presented in section III.\\

From Eq.\ref{eq:BrillianceDivergence1} we found that the quantity that optimizes the x-ray brilliance is
\begin{equation}\label{eq:B1}
     \frac{u^2}{\sigma_e^2}\ \frac{\pi\mathrm{erfcx}(u)^2}{\sqrt{1+  \pi(1+\chi^2)\mathrm{erfcx}(u)^2 }}.
\end{equation}
However, it is possible to set a lower bound on the either the x-ray coherence, determined by $\sigma_e$, or the x-ray energy spread by adding additional parameters to Eq.~\eqref{eq:B1}, resulting in
\begin{equation}\label{eq:B1}
    g(u) =  \frac{u}{1+\sigma_\mathrm{min}^2/u}\ \frac{\pi\mathrm{erfcx}(u)^2}{\sqrt{1+  \pi(1+\chi^2+[\omega_L\sigma_t u \ \varepsilon]^2)\mathrm{erfcx}(u)^2 }},
\end{equation}
where we have used that $u \propto \sigma_e^2$. $\sigma_\mathrm{min}$ is the smallest required source size and $\varepsilon$ is a smallest required x-ray energy spread. $u_\mathrm{max}$ can be numerically evaluated using these additional parameters. The resulting optimization does not optimize the x-ray brilliance, rather the x-ray flux in within a specified partial coherence and energy spread. This can be of interest in for example K-edge subtractive imaging where but high coherence is required spatial resolution, but the energy spread can be $\sim 1-10\%$ \cite{he2012optimization}.\\

The limit where $\sigma_\mathrm{min} \rightarrow \infty$ optimizes the spectral density and $\varepsilon\rightarrow \infty$ optimizes the x-ray brightness. $(\sigma_\mathrm{min},\varepsilon) \rightarrow (\infty,\infty)$ optimizes the x-ray flux.\\

The spectral density is optimized when $u_\mathrm{max} \rightarrow\infty$, whereas the flux and brightness are optimized when $u_\mathrm{max} = 0$. Only when there is a requirement on both the source size and energy spread is the x-ray beam optimized at a finite value of $u_\mathrm{max}$.\\

\section{Comparison with simulations}
Throughout this paper, the developed theory has been checked with simulations. This is done by numerically calculating the spectral angular density from the electron motion with Liénard-Wiechert potentials using the algorithm developed in \cite{thomas2010algorithm}. The electron motion is calculated using a particle tracking simulation (General Particle Tracer \cite{PulsarPhysicsGeneralGPT}) and the electro-magnetic field descriptions of a pulsed Gaussian $\mathrm{TEM}_{00}$ mode with a focus that can be asymmetrical.\\

This simulation not only takes into account a more accurate field description, that includes the radius of curvature, Gouy phase, and a diverging laser pulse, it also includes nonlinear effects and ponderomotive scattering. The electron beam is represented by $10^3$ marcoparticles, where the the effects of electron energy spread, divergence and beam size are incorporated. Since the dimensions of the electron beam are much larger than the radiated wavelength, the contributions from individual microparticles have been added incoherently.

\section{Definition of x-ray brilliance}\label{app:Brilliance}
Throughout this paper we define the x-ray brilliance such that
\begin{equation}
    \int B_\mathrm{x} \  \partial A\ \partial \Omega\  \partial(\omega/\omega_\mathrm{x}) = f N_\mathrm{x},
\end{equation}
according to the convention used in \cite{attwood2000soft}. Note that, using this convention, $B_\mathrm{x}$ is independent on whether r.m.s., FWHM or other values are used to the define the quantities $\Delta A, \Delta\Omega$ and $\Delta\omega/\omega_\mathrm{x}$, as long as a proper normalization is used. Additionally, this definition of x-ray brilliance ensures that Eq.~\ref{eq:Bx_dndwdW} holds. Since $\sigma_e$ and $\Delta\omega/\omega_\mathrm{x}$ refer to r.m.s. quantities, $\Delta A = 2\pi \sigma_e^2$ and  $\Delta\omega/\omega_\mathrm{x} = \sqrt{2\pi} \sigma_\omega/\omega_\mathrm{x}$. Because the x-ray angular spread is uniform, $\Delta \Omega = \pi \Theta_\mathrm{x}^2$. Combining these expressions results in Eq.~\eqref{eq:Brilliance}.

\bibliography{references}

\end{document}